\documentclass[]{interact}

\usepackage{epstopdf}% To incorporate .eps illustrations using PDFLaTeX, etc.
\usepackage{subfigure}% Support for small, `sub' figures and tables

\usepackage{natbib}% Citation support using natbib.sty
\bibpunct[, ]{(}{)}{;}{a}{}{,}% Citation support using natbib.sty
\renewcommand\bibfont{\fontsize{10}{12}\selectfont}% Bibliography support using natbib.sty

\usepackage{url}
\usepackage{lmodern}
\usepackage{bm}
\usepackage{scalerel}

\theoremstyle{plain}% Theorem-like structures provided by amsthm.sty
\newtheorem{theorem}{Theorem}[section]

\newtheorem{corollary}[theorem]{Corollary}
\newtheorem{proposition}[theorem]{Proposition}

\theoremstyle{definition}

\theoremstyle{remark}

\DeclareMathOperator{\expect}{E}

\DeclareMathOperator{\ara}{ARA}
\DeclareMathOperator{\rra}{RRA}
\DeclareMathOperator{\mrs}{MRS}
\DeclareMathOperator{\prs}{PRS}
\DeclareMathOperator{\diag}{diag}
\DeclareMathOperator{\tr}{tr}
\newcommand*{\tran}{^{\mkern-1.5mu\mathsf{T}}}

\begin{document}

\title{Unraveling the Trade-off between Sustainability and Returns: A Multivariate Utility Analysis}

\author{
\name{Marcos Escobar-Anel and Yiyao Jiao \thanks{CONTACT Marcos Escobar-Anel. Email: marcos.escobar@uwo.ca}}
\affil{Department of Statistical and Actuarial Sciences, University of Western Ontario, London, ON, Canada}
}

\maketitle

\begin{abstract}
This paper proposes an expected multivariate utility analysis for ESG investors in which green stocks, brown stocks, and a market index are modeled in a one-factor, CAPM-type structure. This setting allows investors to accommodate their preferences for green investments according to proper risk aversion levels. We find closed-form solutions for optimal allocations, wealth and value functions. As by-products, we first demonstrate that investors do not need to reduce their pecuniary satisfaction in order to increase green investments. Secondly, we propose a parameterization to capture investors' preferences for green assets over brown or market assets, independent of performance. The paper uses the RepRisk Rating of U.S. stocks from 2010 to 2020 to select companies that are representative of various ESG ratings. Our empirical analysis reveals drastic increases in wealth allocation toward high-rated ESG stocks for ESG-sensitive investors; this holds even as the overall level of pecuniary satisfaction is kept unchanged.
\end{abstract}



\section{Introduction}

The concept of environmental, social, and corporate governance (ESG) encompasses a wide range of financial activities, such as sustainable investing, socially responsible investing (SRI), impact investing, green investing, values-based investing, ESG investing, or triple bottom line investing, all of which pointing toward a common objective known as 'green investing'. While each investment style possesses distinct characteristics, they all reflect different aspects of ESG and aim to enhance companies and portfolios in these areas for the benefit of stakeholders. In recent years, numerous institutional investors with long-term investment horizons have integrated elements of ESG into their portfolios. According to a 2022 report by the Sustainable Investment Forum (US SIF, see \citet{sif2022report}), 497 institutional investors, 349 money managers, and 1,359 community investment institutions, which together manage a total of \$7.6 trillion in U.S.-domiciled assets, have incorporated ESG criteria into their investment decision-making and portfolio selection. In other words, ESG is growing in importance  among enterprises, asset managers, and shareholders worldwide \citep{ORSATOISE,edmans2011does,jacobsen2019alpha}. This interest is enabling investors to actively seek investments that generate significant social or environmental benefits, such as community development loan funds or clean technology portfolios. Moreover, the integration of ESG factors is expected to yield long-term financial returns by mitigating potential risks, including those involving litigation, tax, compliance, and reputation \citep{DAM2015,Revelli2015,RIEDL2017}. While the long-term positive impact of prioritizing ESG performance in companies is widely acknowledged, one highly debated question remains: How can investors' preferences for green investing be incorporated into optimal portfolio strategies? This study aims to delve into this precise topic.

One common approach to addressing this question revolves around the inclusion of ethical criteria in addition to the traditional risk-return considerations when optimizing a portfolio \citep{renneboog2008socially}. Studies such as \citet{bollen2007mutual}, \citet{benson2008socially}, and \citet{derwall2011tale} have shown that investors can derive benefits from the sustainable attributes of their investments. In the same spirit, psychological evidence suggests that ESG investors seek both financial returns and non-monetary benefits \citep{pasewark2010sa,webley2001commitment,mackenzie1999morals}, indicating the potential benefits of incorporating additional ESG criteria into portfolio allocation models.

Most authors have incorporated ESG as an ethical, non-pecuniary, dimension into the traditional mean-variance portfolio theory (MVT, see \citet{markowitz1952portfolio}). Some studies such as \citet{chen2021social} screened assets based on ESG criteria before allocating them, while others have included a function of ESG ratings explicitly in the objective function or via allocation constraints \citep{de2021esg,jessen2012optimal,dorfleitner2012theory,pedersen2021responsible,gasser2017markowitz}. When incorporating the ESG dimension into the alternative stream of expected utility theory (EUT), the existing literature predominantly focuses on single-period models. For instance, \citet{ahmed2021modeling} incorporates ESG as a non-pecuniary attribute of the portfolio leading to a bivariate, constant absolute risk aversion (CARA) utility. The study demonstrates that the certainty equivalent (CE) derived from this bivariate utility consistently surpasses the CE obtained from MVT. Similarly, \citet{dorfleitner2017new} work with a bivariate CARA utility where the ESG component appears as a non-pecuniary additive term, and a comparison with MVT is performed. In this setting, they conclude that green investing does not always represent the utility-maximizing solution from an investor's perspective. The celebrated work of \citet{pastor2021sustainable} developed a single-period equilibrium model for green investing with exponential utility. The authors analyzed the impact of ESG preferences on asset prices, suggesting an optimal three-fund separation for portfolio holdings, consisting of a risk-free asset, the market portfolio and an ESG portfolio. 

In parallel with this work, \citet{escobar2022multivariate} incorporates the ESG dimension via a tilting in the investor's risk preferences, i.e., allowing for different risk-aversion levels for green and market assets. This is a pecuniary way of accounting for the ESG ethical dimension. The author also presents the first continuous-time (e.g., continuous rebalancing) ESG analysis in EUT, delivering closed-form solutions for allocations that can be interpreted as a three-fund separation, comprising a risk-free asset, a market portfolio, and a green portfolio. The work also departs from previous literature in its selection of utility functions, employing a constant relative risk aversion (CRRA) utility with two attributes, one for the market portfolio and one for the green
stock.

In our study, we extend the aforementioned paper by proposing a separation of green and brown assets and their corresponding risk preferences. We use a one-factor CAPM setting to model four representative assets: the risk-free asset, a market portfolio, a green stock, and a brown stock. Our analysis decomposes the wealth portfolio into three distinct and independent sources of risk, representing the idiosyncratic green factor, the idiosyncratic brown factor and the systemic (market) factor. We use a multi-attribute constant relative risk aversion (CRRA) utility function to separate the risk preferences of the investor for each of these factors.

This study contributes to the existing literature in several aspects. We list the main points here for clarity:
\begin{itemize}
	\item We perform an expected multivariate utility analysis that enables investors to assign risk-aversion levels to accommodate their preferences for green as opposed to brown investments. The setting permits a closed-form solution for optimal allocations, wealth, and value functions. 
	\item We demonstrate that investors do not need to reduce their overall pecuniary satisfaction in order to increase investments in green stocks. This is achieved via a trade-off between green and brown risk-aversions.
	\item We take this direction further by proposing a parametrization to capture a new dimension of preference, i.e., the investor's inclination toward green assets. By doing so, we connect ESG risk-aversion preferences with ESG ratings. 
	\item The RepRisk rating of U.S. stocks from 2010 to 2020 is used to select green and brown representative companies for empirical analysis. This analysis reveals a critical insight: a recommendation for drastic increases in allocation to green stocks for ESG-sensitive investors.  
\end{itemize}

The remainder of the paper is structured as follows: Section \ref{sec:math} provides the theoretical framework of the paper. Section \ref{sec:optimize} presents its main results and discusses their implications. Section \ref{sec:empirical} 4 provides empirical analysis of examples of the results and their implications, while Section \ref{sec:conclusion} concludes the paper. Appendix \ref{apd:HJB} and \ref{apd:sub} provides proofs for the main results.

\section{Mathematical setting}\label{sec:math}

Let us assume a financial market consisting of one risk-free asset (i.e.,
stock) and three risky assets. Let all the stochastic processes introduced in this paper be defined on a complete probability space $(\Omega ,\mathcal{F},\mathbb{P},\{\mathcal{F%
}_{t}\}_{t\in \lbrack 0,T]})$, where $\{\mathcal{F}_{t}\}_{t\in \lbrack 0, T]} $ is a right-continuous filtration generated by standard Brownian motions (BMs).

This section is divided into three subsections. The first presents the model for the underlying assets and the corresponding synthetic assets. The second subsection constructs the self-financing wealth process and the underlying green, brown, and market synthetic portfolios. The last section introduces the choice of utility, and provides an overview of risk-aversion concepts.

\subsection{ESG market model with a one-factor structure}

We introduce our model, characterized by three main assets. One of the assets is identified as a market portfolio ($S_1$). This could be a common index merging all the assets of a given market or sectors of interest to the investor. The other two assets can be interpreted as two types of stocks in the market: a so-called green stock ($S_2$) and a non-green alternative named, for the purpose of our study, brown stock ($S_3$). Both stocks are correlated with each other and with the index, resembling a one-factor CAPM model, with the index playing the role of the single factor. Our model has the following structure:
\begin{align} \label{ESGmodel}
	\begin{split}
		\frac{dS_{1,t}}{S_{1,t}} &= (r + \lambda_1 \sigma_1^2)dt + \sigma_1 dz_m, \\
		\frac{dS_{2,t}}{S_{2,t}} &= (r + \lambda_1 \sigma_1 \sigma_2 \rho_{12} + \lambda_g \sigma_2^2\sqrt{1-\rho_{12}^2})dt + \sigma_2(\rho_{12} dz_m + \sqrt{1-\rho_{12}^2}dz_g), \\
		\frac{dS_{3,t}}{S_{3,t}} &= (r + \lambda_1 \sigma_1 \sigma_3 \rho_{13} + \lambda_b \sigma_3^2\sqrt{1-\rho_{13}^2})dt + \sigma_3(\rho_{13} dz_m + \sqrt{1-\rho_{13}^2}dz_b).
	\end{split}
\end{align}

where $z_m$, $z_g$, and $z_b$ are independent standard Brownian motions, representing three sources of risk: the market risk, the green risk, and the brown risk respectively. We capture correlations via $corr(S_1, S_2)= \rho_{12}$, $corr(S_1, S_3)= \rho_{13}$, and $corr(S_2, S_3)= \rho_{12}\rho_{13}$. The market risk premium (i.e., from $z_m$) is represented by $\lambda_1 \sigma_1$, the green risk premium is expressed by $\lambda_g \sigma_2$, and the brown risk premium is $\lambda_b \sigma_3$.

It should be noted that the green and brown stocks are driven by market risk and by their corresponding green and brown risks. This means that we can construct synthetic assets governed purely by the green risk or the brown risk; this can be achieved by hedging the market risk as follows:
\begin{align}\label{gbModel}
	\begin{split}
		\frac{dS_{g,t}}{S_{g,t}} &:= \left(\frac{dS_{2,t}}{S_{2,t}} - \frac{dB_t}{B_t}\right)- \beta_2\left(\frac{dS_{1,t}}{S_{1,t}}- \frac{dB_t}{B_t}\right)=\sigma_g(\lambda_g\sigma_2dt + dz_g)\\
		\frac{dS_{b,t}}{S_{b,t}} &:= \left(\frac{dS_{3,t}}{S_{3,t}} - \frac{dB_t}{B_t}\right)- \beta_3\left(\frac{dS_{1,t}}{S_{1,t}}- \frac{dB_t}{B_t}\right)=\sigma_b(\lambda_b\sigma_3dt + dz_b)\\ 
	\end{split}
\end{align}
where $B_t$ denotes the banking process, and the dynamic $dB_t = rB_tdt$, $\beta_2 = \frac{\sigma_2}{\sigma_1}\rho_{12}$, and $\beta_3 = \frac{\sigma_3}{\sigma_1}\rho_{13}$ can be interpreted as the betas of the green and brown stocks in a one-factor structure, while $\sigma_g = \sigma_2\sqrt{1-\rho_{12}^2}$, and $\sigma_b = \sigma_3\sqrt{1-\rho_{13}^2}$ are their non-spanned volatilities. The synthetic assets, $S_g$ and $S_b$ correspond to the exogenous terms in equation (\ref{ESGmodel}). The construction of $S_g$ and $S_b$ will be useful in our later discussion of portfolio optimization.

\subsection{Wealth process and portfolio setting} 

Let $W_t$ denote the investor's wealth process, created by allocating in terms of $S_{1,t}$, $S_{2,t}$, $S_{3,t}$ and $B_t$. Let $\pi_i$ denote the proportion of wealth invested in $S_{i,t}$, according to the self-financing condition:
\begin{align}
	\begin{split}
		\frac{dW_t}{W_t} &= \pi_1\frac{dS_{1,t}}{S_{1,t}} + \pi_2\frac{dS_{2,t}}{S_{2,t}} + \pi_3\frac{dS_{3,t}}{S_{3,t}} + (1-\pi_1-\pi_2-\pi_3)\frac{dB_t}{B_t}\\
		& = (r+\pi_1\lambda_1\sigma_1^2+\pi_2\lambda_1\sigma_1\sigma_2\rho_{12} + \pi_2\lambda_g\sigma_2^2\sqrt{1-\rho_{12}^2}+\pi_3\lambda_1\sigma_1\sigma_3\rho_{13}\\
		&\quad+ \pi_3\lambda_b\sigma_3^2\sqrt{1-\rho_{13}^2})dt 
		+ (\pi_1\sigma_1 + \pi_2\sigma_2\rho_{12}+\pi_3\sigma_3\rho_{13})dz_m+\pi_2\sigma_2\sqrt{1-\rho_{12}^2}dz_g\\
		&\quad+\pi_3\sigma_3\sqrt{1-\rho_{13}^2}dz_b
	\end{split}
\end{align}

The wealth process can be divided into independent components capturing market, green and brown risks. For this purpose, we define the excess return of the market portfolio or index as $\frac{dS_{m,t}}{S_{m,t}} := \frac{dS_{1,t}}{S_{1,t}} - \frac{dB_t}{B_t}$ and $\beta_p := \pi_1 + \pi_2 \beta_2 + \pi_3 \beta_3$. We can alternatively write the wealth process as follows:
\begin{equation}\label{eqn:wealth_process}
	\frac{dW_t}{W_t} = rdt+ \beta_p\frac{dS_{m,t}}{S_{m,t}} + \pi_2\frac{dS_{g,t}}{S_{g,t}} + \pi_3\frac{dS_{b,t}}{S_{b,t}}
\end{equation}
where $S_g$ and $S_b$ are defined in equation (\ref{gbModel}). There is still one extra term in the equation, $rdt$, representing the return of the cash account. We distribute this extra expected return among the three synthetic assets via weighing parameters $\theta_m$, $\theta_g$ and $\theta_b$, satisfying $\theta_m+\theta_g+\theta_b=1$\footnote{For simplicity we will take $\theta_m=1, \theta_g=0, \theta_b=0$ in the numerical section; this has the interpretation of treating the cash account as part of the market portfolio, ideally these weights should be related to the ESG rating of the source of this cash.}. Now we are ready to write the wealth in terms of three synthetic indexes capturing the three independent sources of risk:
\begin{equation*}%\label{WealthandX}
	\frac{dW_t}{W_t} = \frac{dX_{m,t}}{X_{m,t}}+ \frac{dX_{g,t}}{X_{g,t}} +\frac{dX_{b,t}}{X_{b,t}}
\end{equation*}
where
\begin{align}
	\begin{split}
		\frac{dX_{m,t}}{X_{m,t}}&=\beta_p\frac{dS_{m,t}}{S_{m,t}}+\theta_m\frac{dB_t}{B_t} \\
		\frac{dX_{g,t}}{X_{g,t}}&=\pi_2\frac{dS_{g,t}}{S_{g,t}}+\theta_g\frac{dB_t}{B_t}\\
		\frac{dX_{b,t}}{X_{b,t}}&=\pi_3\frac{dS_{b,t}}{S_{b,t}}+\theta_b\frac{dB_t}{B_t}
	\end{split}
\end{align}
We can also write this in the log form:
\begin{equation}\label{WealthandX}
	d\log W_t = d\log X_{m,t}+ d\log X_{g,t}+ d\log X_{b,t}
\end{equation}
where we can see explicitly how each of these synthetic indexes is impacted by its corresponding, independent source of risk:
\begin{align}
	\begin{split}
		d\log X_{m,t} &= [\theta_mr+\pi_1\lambda_1\sigma_1^2+\pi_2\lambda_1\sigma_1\sigma_2\rho_{12}+\pi_3\lambda_1\sigma_1\sigma_3\rho_{13}\\
		&\quad -\frac{1}{2}(\pi_1\sigma_1+\pi_2\sigma_2\rho_{12}+\pi_3\sigma_3\rho_{13})^2]dt + (\pi_1\sigma_1+\pi_2\sigma_2\rho_{12}+\pi_3\sigma_3\rho_{13})dz_m\\
		d\log X_{g,t} &= [\theta_gr+\pi_2\lambda_g\sigma_2^2\sqrt{1-\rho_{12}^2}-\frac{1}{2}\pi_2^2\sigma_2^2(1-\rho_{12}^2)]dt + \pi_2\sigma_2\sqrt{1-\rho_{12}^2}dz_g\\
		d\log X_{b,t} &= [\theta_br+\pi_3\lambda_b\sigma_3^2\sqrt{1-\rho_{13}^2}-\frac{1}{2}\pi_3^2\sigma_3^2(1-\rho_{13}^2)]dt + \pi_3\sigma_3\sqrt{1-\rho_{13}^2}dz_b
	\end{split}
\end{align}

Solving equation (\ref{WealthandX}), we obtain:
\begin{equation}\label{eqn:total_wealth}
	W_T = W_0\frac{X_{m,T}}{X_{m,0}} \frac{X_{g,T}}{X_{g,0}} \frac{X_{b,T}}{X_{b,0}}
\end{equation}

The processes $X_{m,t}$, $X_{g,t}$, and $X_{b,t}$ can be interpreted as indexes denominated in generic units, with their values from the beginning of period 0 to the conclusion of period $T$ dictating the ultimate wealth $W_T$. We refer to $X_{g,t}$ as the Green Index and $X_{b,t}$ as the Brown Index to differentiate them from the green stock ($S_2$) and brown stock ($S_3$) in the market model, as well as distinguishing them from the green synthetic asset ($S_g$) and the brown synthetic asset ($S_b$). Since $X_{m,t}$ is driven by market risk, $X_{g,t}$ by green risk, and $X_{b,t}$ by brown risk, these terms represent their respective contributions to the growth of wealth.

As discussed in the literature (see for instance \cite{escobar2022multivariate}), an investor may prefer to allocate their portfolio according to different degrees of risk aversion for market risk, green risk, and brown risk, respectively. Consequently, we specify the investor's utility as a function of $X_{m,t}$, $X_{g,t}$, and $X_{b,t}$. An investor can maximize their utility by assigning $\pi_1$, $\pi_2$ and $\pi_3$ to get the best combination of $X_{m,t}$, $X_{g,t}$ and $X_{b,t}$. The next section explains the choice of utility.

\subsection{Multivariate utility}
We specify the multivariate utility function as
\begin{equation}\label{ourUtility}
	u(X_m,X_g,X_b) = \frac{(X_m)^{\alpha_m}}{\alpha_m}\frac{(X_g)^{\alpha_g}}{\alpha_g}\frac{(X_b)^{\alpha_b}}{\alpha_b}
\end{equation}
where we choose $\alpha_b\leq\alpha_m\leq\alpha_g<0$. It is evident that the utility function is continuous and twice differentiable. We will explain that our choice of utility satisfies not only the monotonicity of preferences required by many authors, but also various types of risk aversion described in the literature.

\subsubsection{Strictly monotonic preferences}

A decision maker has strictly monotonic preferences over each attribute if the following condition is satisfied:
\begin{equation}
	\frac{\partial u(X_m,X_g,X_b)}{\partial X_i} > 0, i \in \{m,g,b\}
\end{equation}
which indicates that $\alpha_i\alpha_j>0, i \neq j, i \in \{m,g,b\}, j \in \{m,g,b\}$. Thus $\alpha_m < 0, \alpha_g < 0, \alpha_b < 0$ or $\alpha_m > 0, \alpha_g > 0, \alpha_b > 0$ are viable cases.

\subsubsection{KM risk aversion}
As proposed in \citet{km}, a decision maker is KM risk averse if and only if the Hessian of the utility function
\begin{equation}
	\left[ \frac{\partial u(X_1,\ldots,X_n)}{\partial X_i\partial X_j}\right]_{n,n}
\end{equation}
is negative semidefinite. In our case, this means the conditions:
\begin{align}
	\begin{split}
		&\alpha_m < 1, \alpha_g < 1, \alpha_b < 1 \\
		&\alpha_m + \alpha_g < 1, \alpha_m + \alpha_b < 1, \alpha_g + \alpha_b < 1\\
		&\alpha_m + \alpha_g + \alpha_b < 1
	\end{split}
\end{align}
which are all satisfied if $\alpha_b\leq\alpha_m\leq\alpha_g<0$.
\subsubsection{FR risk aversion}
Proposed by \citet{finetti,richard}, FR risk aversion requires that:
\begin{equation}
	\frac{\partial^2 u(\bm{X})}{\partial X_i \partial X_j} \leq 0, \forall i \in \{1,\ldots,N\}, j \in \{1,\ldots,N\}, i \neq j
\end{equation}

Applied to our utility function, this is satisfied as long as
$\alpha_m < 0, \alpha_g < 0, \alpha_b < 0$. 

From \cite{richard}, the condition also guarantees a particular form of utility independence. In our case, $(X_g, X_b) \  u.i.\  X_m$, $X_b\  u.i.\  (X_m, X_g)$, $(X_m, X_g) \  u.i. \  X_b$.

\subsubsection{S risk aversion}
Following the concept of nth-degree risk aversion discussed by \cite{ekern} and \cite{scarsini}, a strictly S risk averse utility function satisfies:
\begin{equation}
	(-1)^{n-1}\frac{\partial^n u(\bm{X})}{\partial X_1 \dots \partial X_n} > 0
\end{equation}

In our case,
\begin{equation}
	(-1)^2 \frac{\partial^3 u(\bm{X})}{\partial X_m \partial X_g \partial X_b} = (X_m)^{\alpha_m}(X_g)^{\alpha_g}(X_b)^{\alpha_b} > 0
\end{equation}
Hence, the utility function in equation (\ref{ourUtility}) is S risk averse as well.

\subsubsection{Summary}
We have shown that for a utility function in the form of equation (\ref{ourUtility}), with $\alpha_i < 0, i \in \{m,g,b\}$, this function has strictly monotonic preferences and is KM-FR-S risk averse. 
%With $\alpha_i > 0, i=1,2,3$ and satisfying equation (2.14), the utility function has strictly monotonic preferences and is KM-S risk averse. 
We can calculate the absolute risk aversion of the utility function with respect to a certain variable, leading to:
\begin{equation}\label{eqn:ara}
	\ara_{X_i}(u) = \frac{1-\alpha_i}{X_i}, i \in \{m,g,b\}.
\end{equation}
indicating that the utility function is in the family of CRRA utility functions, with relative risk aversion being $\rra_{X_i}(u) = 1-\alpha_i$. For $\alpha_m < 0, \alpha_g <0, \alpha_b <0 $, the absolute risk aversion decreases as wealth increases.

Any increase in the return, from any source, would increase the utility of the investor, but we assume that the investor would be more risk-averse to brown risk than green risk, with market risk lying in between. As a result, we introduce our utility function with $\alpha_b\leq\alpha_m\leq\alpha_g<0$.

\section{Optimization problem and solution}\label{sec:optimize}

The objective of an investor is to maximize their utility from terminal wealth. Therefore, the reward function for the investor is defined as follows:
\begin{equation}
	w(X,t;\pi)=\mathbb{E}_{X,t}\left[u(X_{m,T},X_{g,T},X_{b,T})\right] ,  \label{reward function}
\end{equation}%
and the goal is
\begin{equation}\label{ValueFct}
	J(X,t)=\sup_{(\pi)\in \mathcal{U}}w(X,t;\pi),
\end{equation}%
where $J(X,t)$ is the value function and the space $\mathcal{U}$ of
admissible controls $\{\pi _{t}\}_{t\in \lbrack 0,T]}$ with $\pi_{t}\in \mathbb{R}^3$, is the set of feedback strategies that satisfy standard conditions.

As proved in Appendix \ref{apd:HJB_eqn}, for the market model and utility under consideration, we have the following HJB equation:
\begin{equation}\label{eqn:hjb}
	0 = \sup_{\bm{\pi}} \left\{ \nabla_t J + \nabla_{\bm{X}}J\diag (\bm{X_t}) (\bm{r} + \bm{\Lambda} \bm{\lambda}) + \frac{1}{2} \tr (\nabla^2_{\scaleto{\bm{X}\bm{X}}{4pt}}J \diag (\bm{X_t})^2 \bm{\Lambda}^2) \right\}
\end{equation}
where $\bm{\lambda} = [\lambda_1\sigma_1, \lambda_g\sigma_2, \lambda_b\sigma_3]^{\tran}$, $\bm{X_t} = [X_{m,t}, X_{g,t}, X_{b,t}]^{\tran}$, $\bm{r} = [\theta_m, \theta_g, \theta_b]^{\tran}r$, $\bm{\Lambda} = \diag ([\beta_p\sigma_1,\pi_2\sigma_g,\pi_3\sigma_b ])$. The subsequent proposition is obtained through the solution of the HJB equation presented above (see Appendix \ref{apd:HJB_result}).

\begin{proposition}\label{Prop1} The optimal weights solving the HJB equation (\ref{eqn:hjb}) are given by:
	\begin{align}\label{eqn:weights}
		\begin{split}
			\pi_1^* &= \frac{\lambda_1}{1-\alpha_m} - \frac{\sigma_2}{\sigma_1}\rho_{12} \pi_2^* - \frac{\sigma_3}{\sigma_1}\rho_{13} \pi_3^*\\
			\pi_2^* &= \frac{\lambda_g}{\sqrt{1-\rho_{12}^2}(1-\alpha_g)}\\
			\pi_3^* &= \frac{\lambda_b}{\sqrt{1-\rho_{13}^2}(1-\alpha_b)}\\
		\end{split}
	\end{align}
	
	The value function can be expressed in the following form:
	\begin{equation}\label{eq:valueESG}
		J(X_m, X_g, X_b ,t) = \frac{X_m^{\alpha_m}}{\alpha_m}\frac{X_g^{\alpha_g}}{\alpha_g}\frac{X_b^{\alpha_b}}{\alpha_b} \exp (b(T-t))
	\end{equation}
	with $b = \frac{1}{2}\lambda_1^2\sigma_1^2\frac{\alpha_m}{1-\alpha_m} + \frac{1}{2}\lambda_g^2\sigma_2^2\frac{\alpha_g}{1-\alpha_g} + \frac{1}{2}\lambda_b^2\sigma_3^2\frac{\alpha_b}{1-\alpha_b}+(\theta_m\alpha_m+\theta_g\alpha_g+\theta_b\alpha_b)r$.
	
	The optimal wealth process is:
	\begin{equation}
		\frac{dW_t}{W_t} = rdt+ \frac{\lambda_1}{1-\alpha_m}\frac{dS_{m,t}}{S_{m,t}} + \frac{\lambda_g}{\sqrt{1-\rho_{12}^2}(1-\alpha_g)}\frac{dS_{g,t}}{S_{g,t}} + \frac{\lambda_b}{\sqrt{1-\rho_{13}^2}(1-\alpha_b)}\frac{dS_{b,t}}{S_{b,t}}
	\end{equation}
\end{proposition}

It should be noted that an investor's preference for the Green Index does not necessarily mean more allocation to the green stock, the investor may still allocate more to brown stocks. To see this, we can define $m$ as $m = \frac{\lambda_b}{\lambda_g}\sqrt{\frac{1-\rho_{12}^2}{1-\rho_{13}^2}}$. Based on equation (\ref{eqn:weights}), when $m > 1$, we observe that $\pi_3 > \pi_2$ if $\alpha_b > 1-m$ or $\alpha_g < 1-\frac{1}{m}+\frac{\alpha_b}{m}$.

\subsection{A convenient green-brown risk-aversion trade-off}\label{sec:3.1}

In practice, most investors are used to thinking in terms of a general, same-level risk aversion for all sources of risk in the market. Therefore, it makes sense to think of this risk aversion level as $\alpha_m$, i.e., the overall risk aversion to the market portfolio/index. As a way to help practitioners accommodate their potential favoritism toward green stocks, we here provide a simple method to select risk aversion levels for green and brown stocks. With our proposal, there would be no impact on the investor's level of satisfaction arising from past practices, i.e., from the usage of $\alpha_m$ for Green and Brown Indexes. In a nutshell, therefore, we ask the question: What should the risk aversion levels for $\alpha_g$ and $\alpha_b$ be, such that the investor's satisfaction equals the case of $\alpha_g=\alpha_b=\alpha_m$, commonly known as the Merton's solution?

For this purpose, we compare two utility choices for the investor. On the one hand, we assume that the investor follows the popular Merton's utility \citep{merton1969lifetime, merton1975optimum}: $u_M(W_T)=a \frac{W_T^{\alpha_m}}{\alpha_m}$, where the scalar $a$ will be specified later. In this case, the investor assigns the same degree of risk-aversion to $X_m$, $X_g$, and $X_b$. To see this, from equation (\ref{eqn:total_wealth}), define $c= \frac{W_0}{X_{m,0}X_{g,0}X_{b,0}}$, then $u_M(W_T) = ac^{\alpha_m}\frac{(X_{m,T}X_{g,T}X_{b,T})^{\alpha_m}}{\alpha_m}$. 

On the other hand, the investor uses our multivariate utility $u(W_T)=\frac{X_{m,T}^{\alpha_m}}{\alpha_m} \frac{X_{g,T}^{\alpha_g}}{\alpha_g}\frac{X_{b,T}^{\alpha_b}}{\alpha_b}$, where $\alpha_b \leq \alpha_m \leq \alpha_g <0$. In order to ensure a fair comparison of the two utilities, we assume $u_M(W_0)=u(W_0)$, which means that $a=\frac{X_{m,0}^{\alpha_m}X_{g,0}^{\alpha_g}X_{b,0}^{\alpha_b}}{W_{0}^{\alpha_m}\alpha_g \alpha_b}$. 

Let us define the value function for the Merton solution as $J_M(W, t)$, and recall that the value function for the ESG utility is $J(X, t)$. Our objective is to determine a set of risk aversions $\alpha_g$ and $\alpha_b$, with $\alpha_b \leq \alpha_m \leq \alpha_g <0$, such that $J_M=J$. In other words, the investor achieves the same level of satisfaction while accommodating less risk aversion to the Green Index $X_g$ and more risk aversion to the Brown Index $X_b$.

The Merton solution can be easily derived from our approach  by setting the risk aversion levels properly, i.e., having $\alpha_g=\alpha_b=\alpha_m$ in equation (\ref{eq:valueESG}). This leads to:
\begin{equation}\label{eqn:merton sol}
	J_M(W_t, t) = a\frac{W_t^{\alpha_m}}{\alpha_m} \exp (b_M(T-t))
\end{equation}
where $b_M = \frac{1}{2}(\lambda_1^2\sigma_1^2 + \lambda_g^2\sigma_2^2+\lambda_b^2\sigma_3^2 )\frac{\alpha_m}{1-\alpha_m} + \alpha_mr$. 
The relation $J_M=J$ can be expressed as follows:
\begin{equation}
	e^{(b_M - b)(T-t)} = \left(\frac{X_{g,t}}{X_{g,0}}\right)^{\alpha_g - \alpha_m}\left(\frac{X_{b,t}}{X_{b,0}}\right)^{\alpha_b - \alpha_m}
\end{equation} 
Now we are ready for the result.

\begin{corollary}\label{Coro1} 
	At $t=0$ and assuming $\theta_m=1, \theta_g=\theta_b=0$, the following relation between risk aversion levels ensures $J_M=J$:
	\begin{equation}\label{eqn:optimalTradeoff}
		\frac{\alpha_g}{1-\alpha_g} =( 1+\frac{\lambda_b^2\sigma_3^2}{\lambda_g^2\sigma_2^2} )\frac{\alpha_m}{1-\alpha_m}  - \frac{\lambda_b^2\sigma_3^2}{\lambda_g^2\sigma_2^2}\frac{\alpha_b}{1-\alpha_b}
	\end{equation}
\end{corollary}
The proof follows easily from equation $J_M=J$. 

This relation can be interpreted as follows: for any investor’s given level of risk-aversion, i.e. $\alpha_m$, there is a family of pairs $(\alpha_g, \alpha_b)$ determined by equation (\ref{eqn:optimalTradeoff}), such that the investor can achieve the dual purpose of maintaining the same level of satisfaction as with the (targeted, classical, existing) Merton's investment while ensuring their ESG risk-preference for the green stock ($\alpha_g$) is fulfilled. Naturally, a greater preference for green ($\alpha_g>\alpha_m$) would come at the expense of less preference for brown ($ \alpha_b < \alpha_m$).

\subsection{A parameterization of ESG risk aversion levels}\label{sec:3.2}

The previous section shows the possibility of a trade-off between green and brown preferences. This means that the investor can increase their risk aversion to brown and decrease the risk aversion to green while keeping the value function, satisfaction, unchanged. This trade-off is captured by the relation between $\alpha_g$ and $\alpha_b$ in equation (\ref{eqn:optimalTradeoff}). Unfortunately, such a relationship depends on the performance of the underlying assets (e.g., $\lambda_2$, $\sigma_3$, etc). For an investor, the preference for green over brown shall be independent of the stock price performance and fully based on the ratings of the stock.

In this spirit, let us assume that an investor adjusts their risk aversion by comparing the ESG rating of a company with that of the market. Consider a discrete mapping from ESG ratings to ESG scores, where ESG scores are positive integers denoted by $E$, e.g. from 1 (D) to 10 (AAA), hence the larger the $E$ the 'greener' the ESG rating. We propose to model the risk aversion to green and brown stocks as follows: $\alpha_i = \alpha_m \exp{(\kappa (E_m - E_i))}$ with $\kappa > 0$, $i \in \{g, b\}$, where $E_m$ denotes the ESG score of the market portfolio. This model ensures that $\alpha_g \leq \alpha_m \leq \alpha_b < 0$.

The parameter $\kappa$ can be interpreted in two ways. First, for all $X_i$, we can write
\begin{equation}
	\kappa = \frac{1}{1-\rra_{X_i}}\frac{\partial \rra_{X_i}}{\partial E_i}
\end{equation}
which indicates that $\kappa$ determines relative decreases in $\rra_{X_i}$ due to increases in the ESG rating/score $E_i$. Investors with a value of $\kappa$ represent people who are more sensitive in their risk aversion levels to the ESG ratings.

Second, $\kappa$ also affects the substitution effect in utility between different values of $X_i$. To see this, let $\mathrm{H}$ denote $\{m,g,b\}$, with the marginal rate of substitution (MRS) given as:
\begin{equation}
	\mrs_{ij} = \frac{U_{X_i}}{U_{X_j}}=\frac{\alpha_i X_j}{\alpha_j X_i}=\exp{(\kappa (E_j - E_i))}\frac{X_j}{X_i}
\end{equation}
where $i \in \mathrm{H}$, $j \in \mathrm{H}$, $i \neq j$. Furthermore, fixing $X_k$, $k \in \mathrm{H}$, we have $U_{X_i}dX_i + U_{X_j}dX_j = dU = 0$, $i \in \mathrm{H}$, $j \in \mathrm{H}$, $i \neq j \neq k$. We can calculate the percentage rate of substitution (PRS) among $X_m, X_g, X_b$:
\begin{equation}
	\prs_{ij} = -\frac{dX_i/X_i}{dX_j/X_j} = \frac{U_{X_j}}{U_{X_i}}\frac{X_j}{X_i}=\frac{\alpha_j}{\alpha_i}=\exp{(\kappa (E_i - E_j))}
\end{equation}
where $i \in \mathrm{H}$, $j \in \mathrm{H}$, $i \neq j$. This leads to the following representations for $\kappa$
\begin{equation}
	\kappa = -\frac{\frac{d \mrs_{ij}}{\mrs_{ij}}}{d(E_i-E_j)} = \frac{\frac{d \prs_{ij}}{\prs_{ij}}}{d(E_i-E_j)}
\end{equation}
Hence, $\kappa$ captures the degree of influence that a change in ESG rating has on a relative change in $\mrs$ and $\prs$.

To gain a deeper understanding of this result, let us interpret $\prs_{ij}$. We observe that a 1\% increase in the quantity of good $X_j$ is equivalent, from the investor's perspective, to a $\prs_{ij}$\% decrease in the quantity of good $X_i$. Assuming $E_b < E_m < E_g$, we find that $\prs_{bg} < 1$, $\prs_{mg} < 1$, and $\prs_{mb} > 1$. As the value of $\kappa$ increases, both $\prs_{bg}$ and $\prs_{mg}$ decrease, while $\prs_{mb}$ increases. This suggests that as $\kappa$ increases, investors become more inclined to enhance their green wealth with relatively less emphasis on brown or market wealth.

\subsection{Suboptimal strategies and wealth equivalent losses.}\label{sec:3.3}

This section aims to investigate the wealth equivalent losses (WEL) that investors may experience if they use suboptimal strategies. An investor might adopt a suboptimal strategy for many reasons. One common rationale is a lack of knowledge about building an optimal strategy. For instance, in our setting, an investor might be inclined to assign different degrees of risk aversion to their Green and Brown Indexes but they lack the knowledge to craft an optimal solution; hence, they decide to use the well-known Merton's solution. The resulting allocations for the investor, determined using the same risk aversion level, would be suboptimal, leading to a decrease in utility. 

In this study, we denote the value function obtained from a suboptimal strategy $\bm{\pi}^s$ as $J^s$, and we define the Green-Index Wealth Equivalent Loss (GWEL) as the scalar $q$, satisfying the following equation:
\begin{equation}\label{eqn:cel}
	J(X_m, X_g(1-q), X_b, 0) = J^s(X_m, X_g, X_b, 0)
\end{equation}

It should be noted that GWEL is very similar to the conventional definition of WEL, where WEL is the $q$ such that $J(W(1-q), 0) = J^s(W, 0)$. In our case, the parameter $q$ represents the percentage-wise reduction in the value of the Green Index that the optimal investor can incur while performing at the same level of satisfaction as the suboptimal investor. This interpretation directly captures the extent to which the optimal portfolio would lose its 'greenness' to mimic suboptimal allocation by investors.

Let us choose Merton's solution in equation (\ref{eqn:merton sol}) as our suboptimal strategy and denote WEL as $q_m$, where:
\begin{align}
	\begin{split}
		\pi_1^s &= \frac{\lambda_1}{1-\alpha_m} - \frac{\sigma_2}{\sigma_1}\rho_{12} \pi_2^s - \frac{\sigma_3}{\sigma_1}\rho_{13} \pi_3^s\\
		\pi_2^s &= \frac{\lambda_g}{\sqrt{1-\rho_{12}^2}(1-\alpha_m)}\\
		\pi_3^s &= \frac{\lambda_b}{\sqrt{1-\rho_{13}^2}(1-\alpha_m)}\\
	\end{split}
\end{align}

For the suboptimal strategy $\bm{\pi}^s$, the value function would be (see Appendix \ref{apd:sub_merton} for derivations):
\begin{equation}
	J^s(X_m, X_g, X_b ,t) = \frac{X_m^{\alpha_m}}{\alpha_m}\frac{X_g^{\alpha_g}}{\alpha_g}\frac{X_b^{\alpha_b}}{\alpha_b} \exp (b^s(T-t))
\end{equation}
where $b^s = \frac{1}{2}\lambda_1^2\sigma_1^2\frac{\alpha_m}{1-\alpha_m}+\lambda_g^2\sigma_2^2\frac{\alpha_g}{1-\alpha_m}-\frac{1}{2}\lambda_g^2\sigma_2^2\frac{\alpha_g(1-\alpha_g)}{(1-\alpha_m)^2}+\lambda_b^2\sigma_3^2\frac{\alpha_b}{1-\alpha_m}-\frac{1}{2}\lambda_b^2\sigma_3^2\frac{\alpha_b(1-\alpha_b)}{(1-\alpha_m)^2}+(\theta_m\alpha_m+\theta_g\alpha_g+\theta_b\alpha_b)r$.

Resorting to equations (\ref{eq:valueESG}) and (\ref{eqn:cel}), we obtain:
\begin{equation}\label{eqn:subopt}
	e^{(b^s-b)T}=(1-q_m)^{\alpha_g}
\end{equation}
This is:
\begin{equation}\label{eqn:wel}
	q_m=1-\exp\left\{\frac{(b^s-b)T}{\alpha_g}\right\}
\end{equation}

We are also interested in the WEL for an investor choosing not to invest in the green stock, referred to as $q_g$. Solving the HJB equation (\ref{eqn:hjb}) while fixing $\pi_2^s = 0$ leads to the following allocations:
\begin{align}
	\begin{split}
		\pi_1^s &= \frac{\lambda_1}{1-\alpha_m}  - \frac{\sigma_3}{\sigma_1}\rho_{13} \frac{\lambda_b}{\sqrt{1-\rho_{13}^2}(1-\alpha_b)}\\
		\pi_2^s &= 0\\
		\pi_3^s &= \frac{\lambda_b}{\sqrt{1-\rho_{13}^2}(1-\alpha_b)}\\
	\end{split}
\end{align}

The value function is (see Appendix \ref{apd:sub_nongreen} for derivations):
\begin{equation}
	J^s(X_m, X_g, X_b ,t) = \frac{X_m^{\alpha_m}}{\alpha_m}\frac{X_g^{\alpha_g}}{\alpha_g}\frac{X_b^{\alpha_b}}{\alpha_b} \exp (b^s(T-t))
\end{equation}
with $b^s = \frac{1}{2}\lambda_1^2\sigma_1^2\frac{\alpha_m}{1-\alpha_m} + \frac{1}{2}\lambda_b^2\sigma_3^2\frac{\alpha_b}{1-\alpha_b}+(\theta_m\alpha_m+\theta_g\alpha_g+\theta_b\alpha_b)r$. Similarly, we have:
\begin{equation}\label{eqn:wel_2}
	q_g = 1-\exp\left\{-\frac{1}{2}\lambda_g^2\sigma_2^2\frac{T}{1-\alpha_g}\right\}
\end{equation}

\section{Empirical results}\label{sec:empirical}

We work with ESG ratings as obtained from the RepRisk database. By mapping the RepRisk Rating to integers from 1 (D) to 10 (AAA) and then computing the average RRR score, we identify the top 10 U.S. companies as green companies and the bottom 10 as brown companies (refer to Table \ref{tab:us20}). The ESG rating for the U.S. market portfolio is determined by calculating the mean ESG rating across all U.S. companies, resulting in a value of 7.3.

\begin{table}
	\tbl{Top 10 green companies and bottom 10 brown companies in the US}
	{\begin{tabular}{lr}
			\hline
			Company name                                   & \multicolumn{1}{r}{Averaged ESG rating} \\ \hline
			Andalay Solar Inc                              & 9.393939394                \\
			Clearway Energy Inc (formerly NRG Yield   Inc) & 9.393939394                \\
			Ocean Power Technologies Inc                   & 9.393939394                \\
			BellSouth Telecommunications Inc   (formerly South Central Bell Telephone Co)    & 9.393939394 \\
			RCN Telecom Services LLC                       & 9.386363636                \\
			Alltel Corp                                    & 9.386363636                \\
			IDT Corp                                       & 9.386363636                \\
			Shenandoah Telecommunications Co   (Shentel)   & 9.378787879                \\
			Maidenform Brands Inc                          & 9.356060606                \\
			Friedman's Inc                                 & 9.356060606                \\
			Bunge Ltd (formerly Bunge \& Born)             & 4.340909091                \\
			Inter-American Development Bank (Banco   Interamericano de Desarrollo; BID; IDB) & 4.295454545 \\
			Wells Fargo \& Co                              & 4.234848485                \\
			EI du Pont de Nemours \& Co (DuPont)           & 4.212121212                \\
			Target Corp                                    & 4.015151515                \\
			Apple Inc (Apple)                              & 3.992424242                \\
			Bank of America Corp (BOA)                     & 3.871212121                \\
			Cargill Inc                                    & 3.560606061                \\
			Walmart Inc (formerly Wal-Mart Stores   Inc; Walmart)                            & 3.431818182 \\
			World Bank Group (WBG); The                    & 2.568181818       \\ \hline        
		\end{tabular}}
	\label{tab:us20}
\end{table}

For our market process, we select the S\&P 500 as our index. Considering the availability of stock data, we choose two pairs of stocks: the first pair includes IDT Corp. (IDT), representing the green company, and Walmart Inc. (WMT), representing the brown company; the second pair is Shenandoah Telecommunications Company (Shenandoah) as the green company and DuPont de Nemours, Inc. (DuPont) as the brown company. Table \ref{tab:parameters} presents the parameters for our model in these two cases, estimated on a monthly basis. In this section, we present empirical findings for two distinct configurations: the first involving IDT and WMT, the second incorporating Shenandoah and DuPont. The risk-free rate is calculated by averaging the yields of the 3-month Treasury bill issued by the U.S. government.

\begin{table}
	\tbl{Parameter estimations for empirical analysis}
	{\begin{tabular}{lrrrr}
		\hline
		Name & $\sigma$ & $\rho$ & $\lambda$ & ESG rating \\ \hline
		S\&P 500             & 0.0405   & 1      & 6.0464    & 7.3        \\
		IDT                  & 0.1628   & 0.2937 & 0.7       & 9.4        \\
		Walmart              & 0.0486   & 0.3354 & 2.8672    & 3.4        \\
		Shenandoah Telecom   & 0.1064   & 0.291  & 1.0179    & 9.4        \\
		DuPont de Nemours    & 0.0866   & 0.767  & -1.244    & 4.2        \\ \hline
	\end{tabular}}
	\label{tab:parameters}
\end{table}

The analysis is divided into three parts, along the lines of Section \ref{sec:optimize}. In Section \ref{sec:impact}, we illustrate the solution implied by Proposition \ref{Prop1}, which is the impact of different risk aversions for Green and Brown Indexes on optimal allocations. In Section \ref{sec:ea_tradeoff}, we focus on exposing the trade-off region $(\alpha_g, \alpha_b)$ according to Corollary \ref{Coro1}. Section \ref{sec:ea_para} provides insights into the impact of the parametrization of risk-aversions motivated in Section \ref{sec:3.2}, while Section \ref{sec:ea_gwel} focuses on the GWEL analysis explained in Section \ref{sec:3.3}.

\subsection{Impact of green and brown risk aversions}\label{sec:impact}

\begin{figure}
	\makebox[\textwidth][c]{\includegraphics[width=1.14\textwidth]{"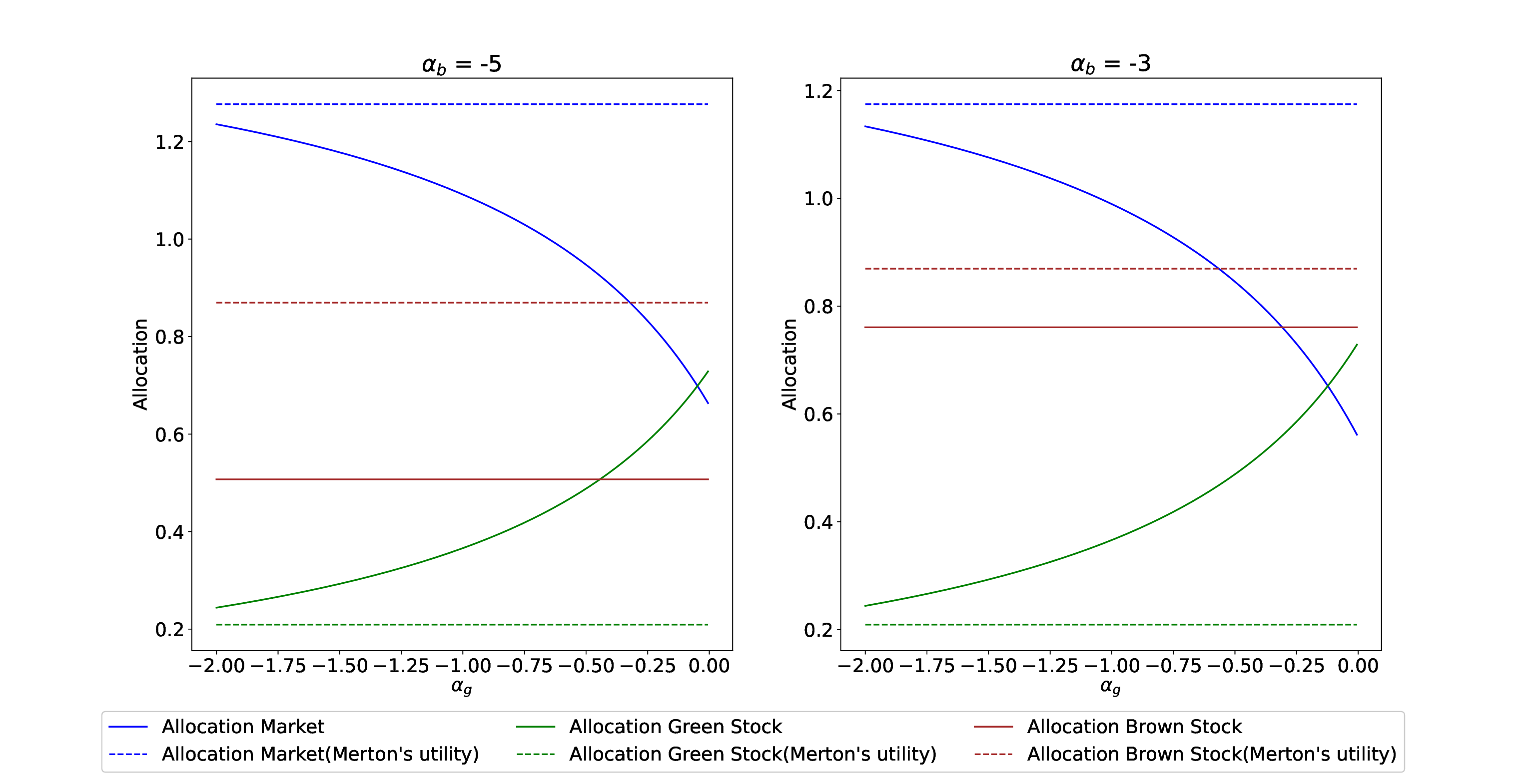"}}
	\caption{Impact of difference in green risk aversion, IDT vs. WMT, $\alpha_m$ = -2.5"}
	\label{fig:chg_a2}
\end{figure}

\begin{figure}
	\makebox[\textwidth][c]{\includegraphics[width=1.14\textwidth]{"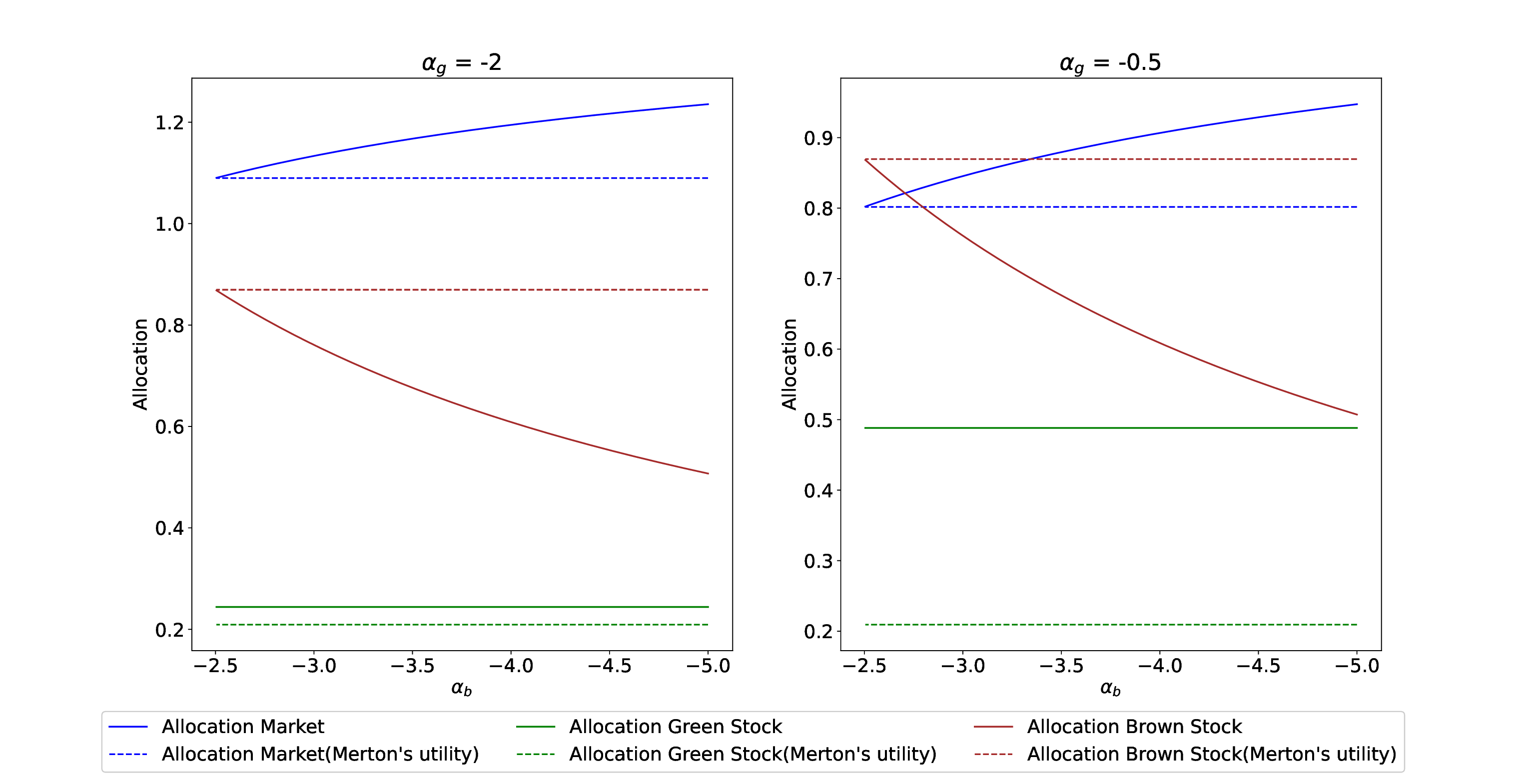"}}
	\caption{Impact of difference in brown risk aversion, IDT vs. WMT, $\alpha_m$ = -2.5}
	\label{fig:chg_a3}
\end{figure}

For Merton's allocation approach, we take $\alpha_m = \alpha_g = \alpha_b = -2.5$ as the benchmark represented by dotted lines in the figures. Figure \ref{fig:chg_a2} demonstrates the impact of increasing $\alpha_g$ from -2 to 0, while keeping $\alpha_b=-5$, as well as $\alpha_b=-3$. Figure \ref{fig:chg_a3} highlights the effect of increasing risk aversion toward Brown Index, i.e., decreasing $\alpha_b$ from -2.5 to -5, while keeping $\alpha_g$ constant at -2 in the first graph and -0.5 in the second graph. A comparison between these two figures clearly reveals a substantial increase and decrease in the allocation to green and brown stocks, respectively. The actual behavior is not surprising; the interesting aspect is the size of the changes in allocation. For example, in Figure \ref{fig:chg_a2}, the allocation to the green stock goes from roughly 20\% to close to 70\%, all of this change resulting in a reduction of the allocation to the market portfolio.

\begin{figure}
	\makebox[\textwidth][c]{\includegraphics[width=1.14\textwidth]{"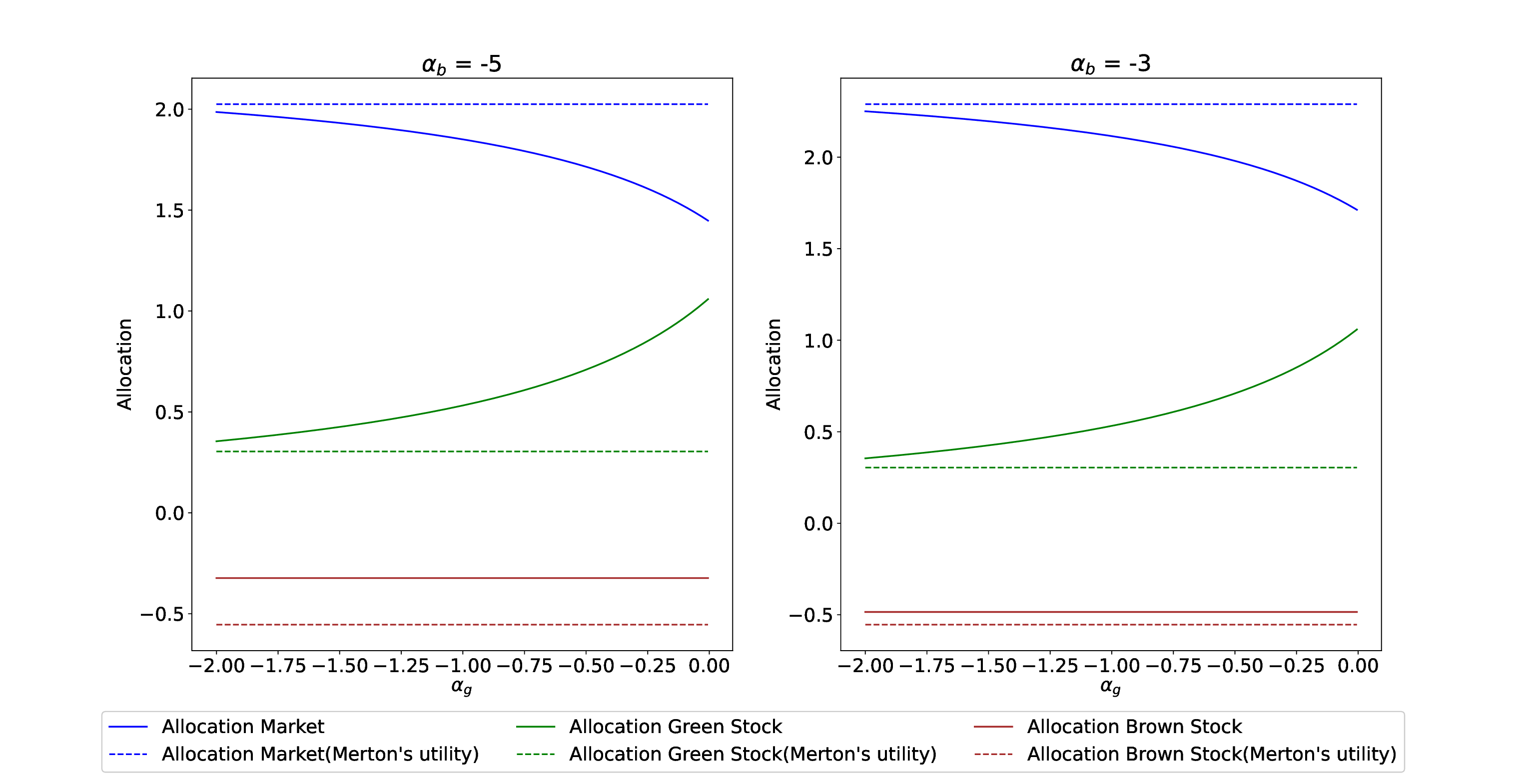"}}
	\caption{Impact of difference in green risk aversion, Shenandoah vs. DuPont, $\alpha_m$ = -2.5}
	\label{fig:chg_a2_sd}
\end{figure}

\begin{figure}
	\makebox[\textwidth][c]{\includegraphics[width=1.14\textwidth]{"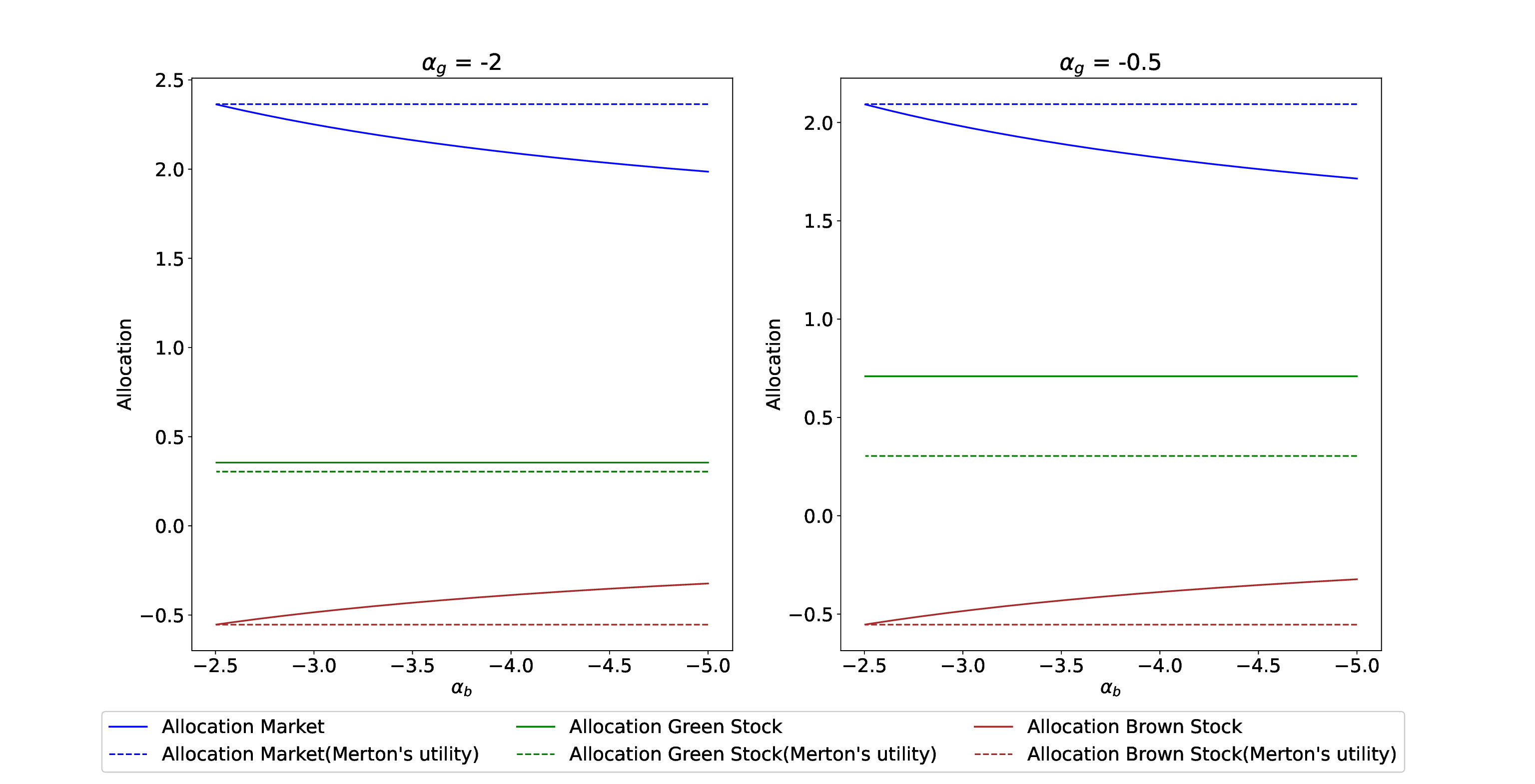"}}
	\caption{Impact of difference in brown risk aversion, Shenandoah vs. DuPont, $\alpha_m$ = -2.5}
	\label{fig:chg_a3_sd}
\end{figure}

This same analysis was performed in the second pair of stocks, Shenandoah and DuPont, in Figures \ref{fig:chg_a2_sd} and \ref{fig:chg_a3_sd}, with similar results. Both figures demonstrate a negative weight assigned to the brown stock. According to Equation (\ref{eqn:weights}), this indicates that when $\lambda_b < 0$, investors engage in short selling of the brown stock, which in this specific case is DuPont. Nevertheless, in such a scenario, a higher level of risk aversion towards the Brown Index would imply a decrease in the amount of brown stocks being short sold.

Next, we aim to explore the circumstances under which an investor would allocate more to the green stock. As stated in Section \ref{sec:3.1}, if $m\leq1$ and $\alpha_b\leq\alpha_m\leq\alpha_g<0$, the investment in the green stock exceeds that in the brown stock. However, for $m > 1$, specific parameter combinations must strictly satisfy the condition $\alpha_b < 1-m$ and $\alpha_g > 1-\frac{1}{m}+\frac{\alpha_b}{m}$ in order for investors to prioritize the green stock over the brown stock. In the case of IDT and WMT, where $m = 4.16 > 1$, we find that $\alpha_b < -3.1561$ and $\alpha_g > 0.7594 + 0.2406\alpha_b$, as long as $\alpha_b \leq \alpha_m \leq \alpha_g < 0$. Figure \ref{fig:chg_a2} confirms that when $\alpha_b = -3$, it is not possible for the weight of the green stock to exceed the weight of the brown stock. Additionally, when $\alpha_b = -5$, $\alpha_g > -0.4423$ is required to ensure greater investment in the green than the brown stock. The same analysis is repeated for Shenandoah and DuPont, resulting in $m = -1.8222 \leq 1$, which guarantees that the weight of the green stock consistently surpasses the weight of the brown.

\subsection{Green-brown risk-aversion trade-off analysis}\label{sec:ea_tradeoff}

\begin{figure}
	\makebox[\textwidth][c]{\includegraphics[width=1.15\textwidth]{"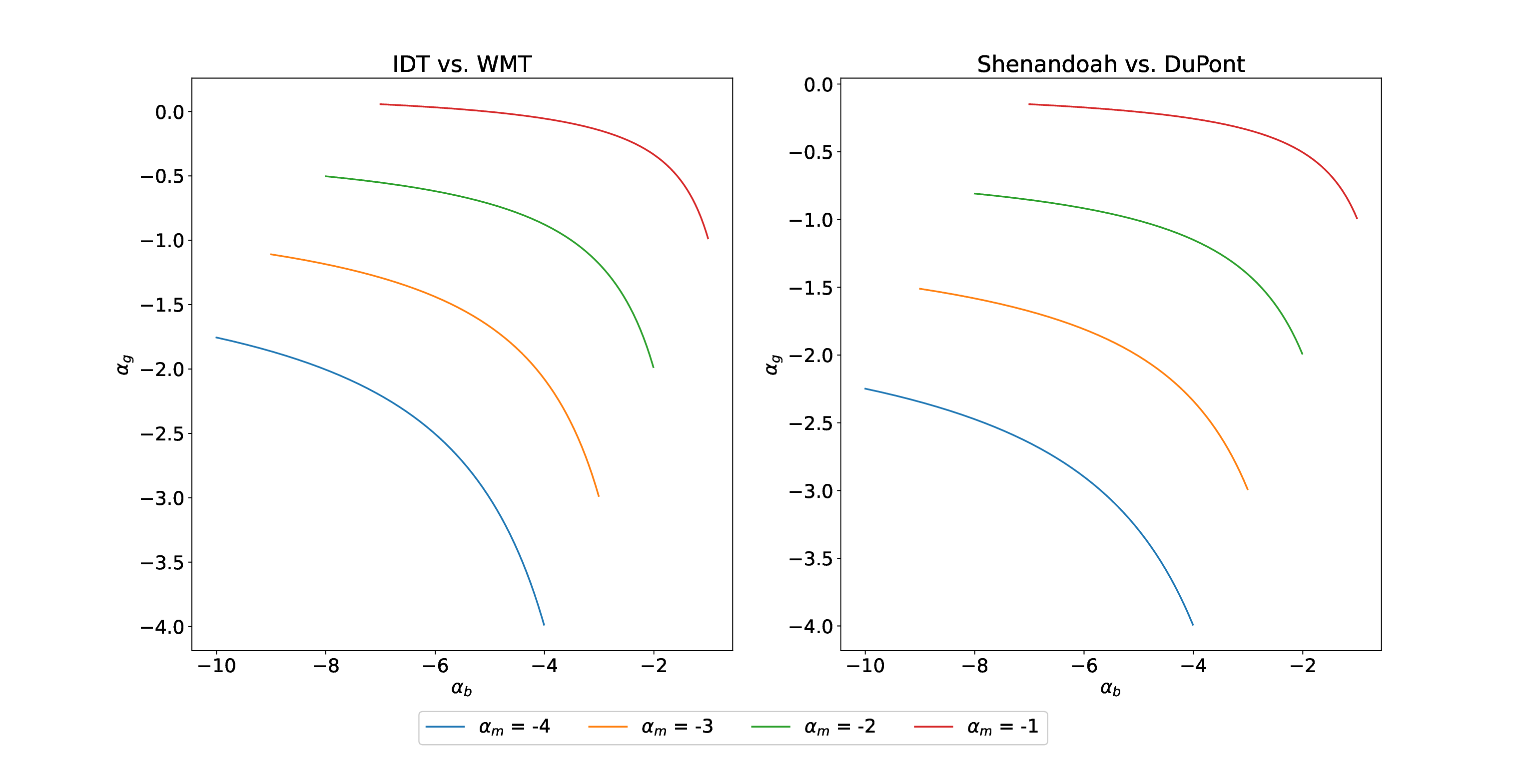"}}
	\caption{Green-Brown risk-aversion trade-off}
	\label{fig:tradeoff}
\end{figure}

In Section \ref{sec:3.1}, we discuss the set of pairs $(\alpha_g, \alpha_b)$ that enable the preservation of Merton's utility level while incorporating ESG risk preferences. As depicted in Figure \ref{fig:tradeoff} (using equation (\ref{eqn:optimalTradeoff})), investors can uphold their utility by appropriately calibrating their risk aversion. The graph illustrates that investors have the freedom to express their ESG preferences without experiencing a compulsory decline in utility levels, even in diverse market conditions.

In particular, the figure shows, for a given level of $\alpha_m$, the curve of pairs  $(\alpha_g, \alpha_b)$ leading to the same satisfaction level (value function $J$). The lower the market risk aversion level the shorter the regions for the pairs. Nevertheless, there is plenty of flexibility to select $(\alpha_g, \alpha_b)$; every pair within the curve describes a different investor, and those toward the northwest of the figure can be interpreted as more aggressive to the Green Index. This level of flexibility gives rise to the notion of a parametrization or a measure to assess the ESG risk preference, as explained in Section \ref{sec:3.2} and illustrated in Section \ref{sec:ea_para} below. 

\begin{figure}
	\makebox[\textwidth][c]{\includegraphics[width=1.14\textwidth]{"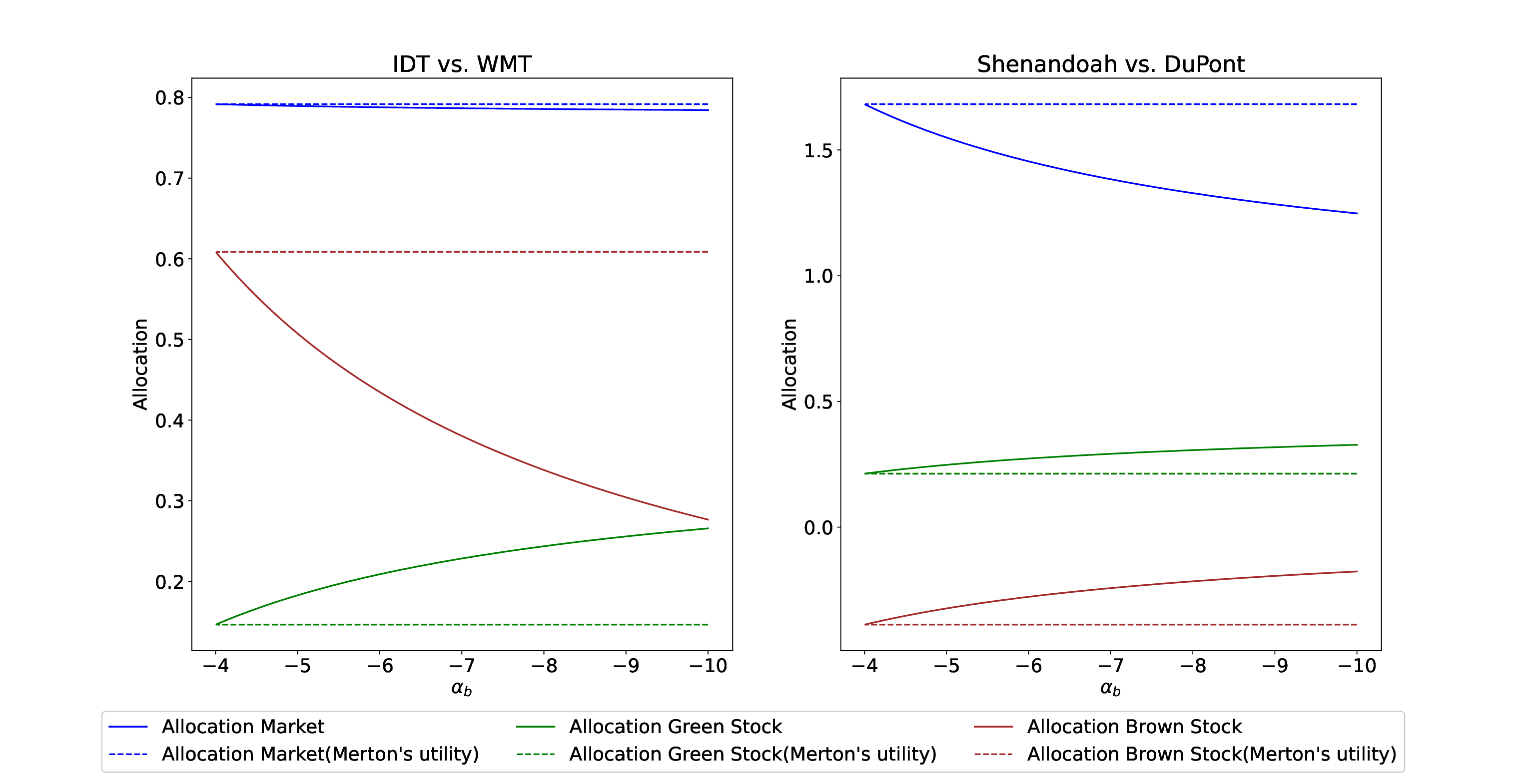"}}
	\caption{Impact of difference in brown risk aversion under Green-Brown risk-aversion trade-off, $\alpha_m = -4$}
	\label{fig:tradeoff_alloc}
\end{figure}

It is reasonable to inquire about the impact of ESG preferences on allocations while maintaining the same level of satisfaction as Merton's utility. To illustrate this, let us consider $\alpha_m = -4$. Figure \ref{fig:tradeoff_alloc} demonstrates the variations in portfolio weights as we traverse along the blue curve depicted in Figure \ref{fig:tradeoff}. The figure reveals that investors with a preference for green assets would allocate a significantly greater proportion of their portfolio to green assets while greatly reducing their brown allocation, without compromising their utility levels. This flexibility allows diverse investors to allocate their portfolios in a manner that is simple and aligns with their ESG preferences.

\subsection{Studying the parametrization of ESG risk-aversion levels}\label{sec:ea_para}

In Section \ref{sec:3.2}, we introduced the parameter $\kappa$ to capture ESG risk preferences. Taking $\alpha_i = \alpha_m e^{\kappa (E_m - E_i)}, i \in \{g, b\}$ into equation (\ref{ourUtility}), we have:
\begin{equation}
	X_g^{p_1(\kappa)}X_b^{p_2(\kappa)} = \frac{((\alpha_m \alpha_g \alpha_b)u(X_m,X_g,X_b))^{1/\alpha_m}}{X_m}=C
\end{equation}
where $p_1(\kappa) = e^{\kappa(E_M-E_2)}$ and $p_2(\kappa) = e^{\kappa(E_M-E_3)}$. Note $C$ is a constant if assuming a fixed utility level and fixed Market Index $X_m$.

\begin{figure}
	\makebox[\textwidth][c]{\includegraphics[width=1.2\textwidth]{"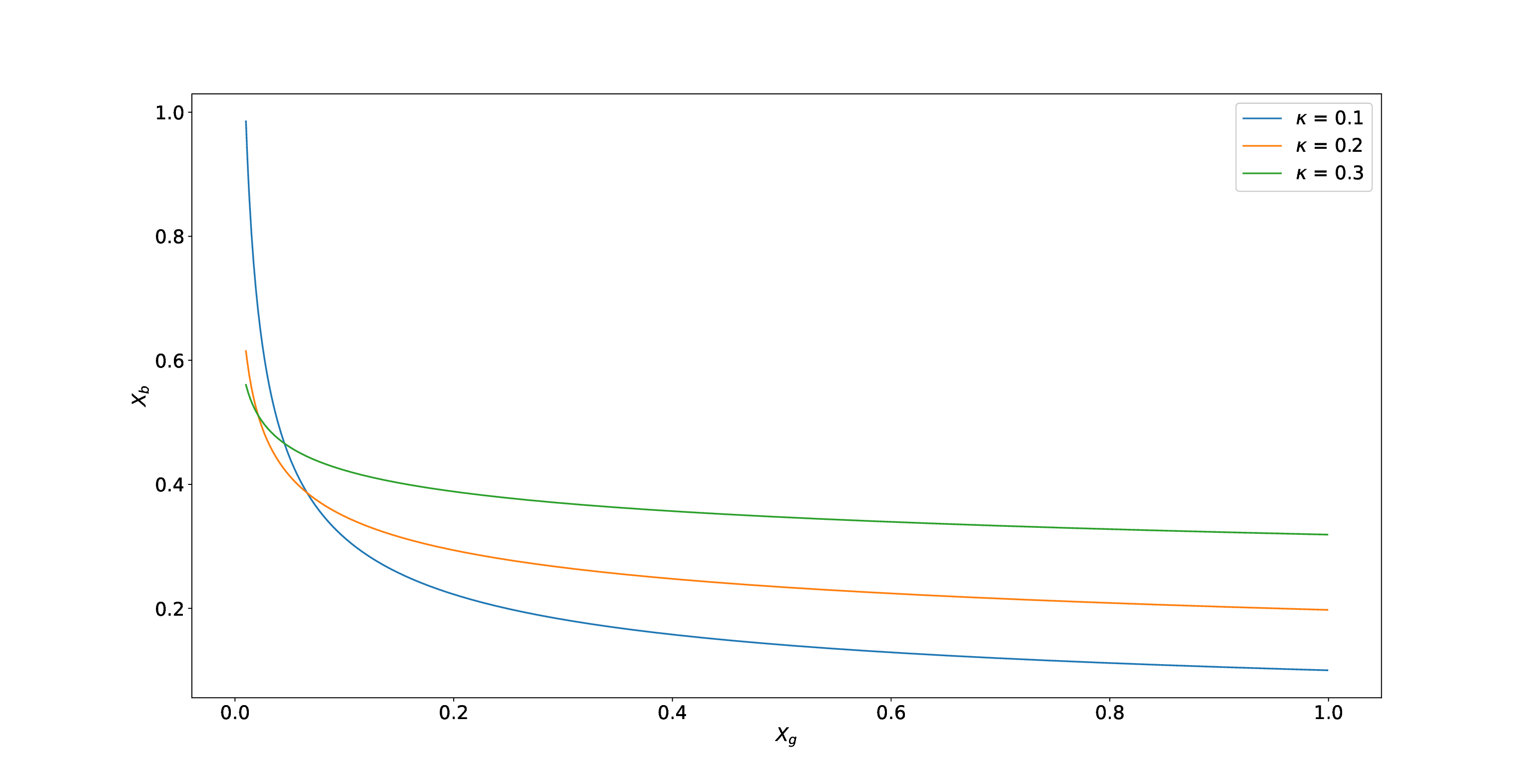"}}
	\caption{Indifference curve for the same utility level with different \(\kappa\)}
	\label{fig:indiff}
\end{figure}

Figure \ref{fig:indiff} illustrates how $\kappa$ influences the indifference curve between $X_g$ and $X_b$ in the equation above at a specific utility level. It demonstrates a flatter curve favoring Green Index and a decrease in $\mrs_{bg}$ and $\prs_{bg}$ as $\kappa$ increases. This empirical observation supports the idea that higher values of $\kappa$ facilitate investor mobility toward a larger allocation to Green Index while experiencing relatively smaller reductions in Brown Index.

\begin{figure}
	\makebox[\textwidth][c]{\includegraphics[width=1.14\textwidth]{"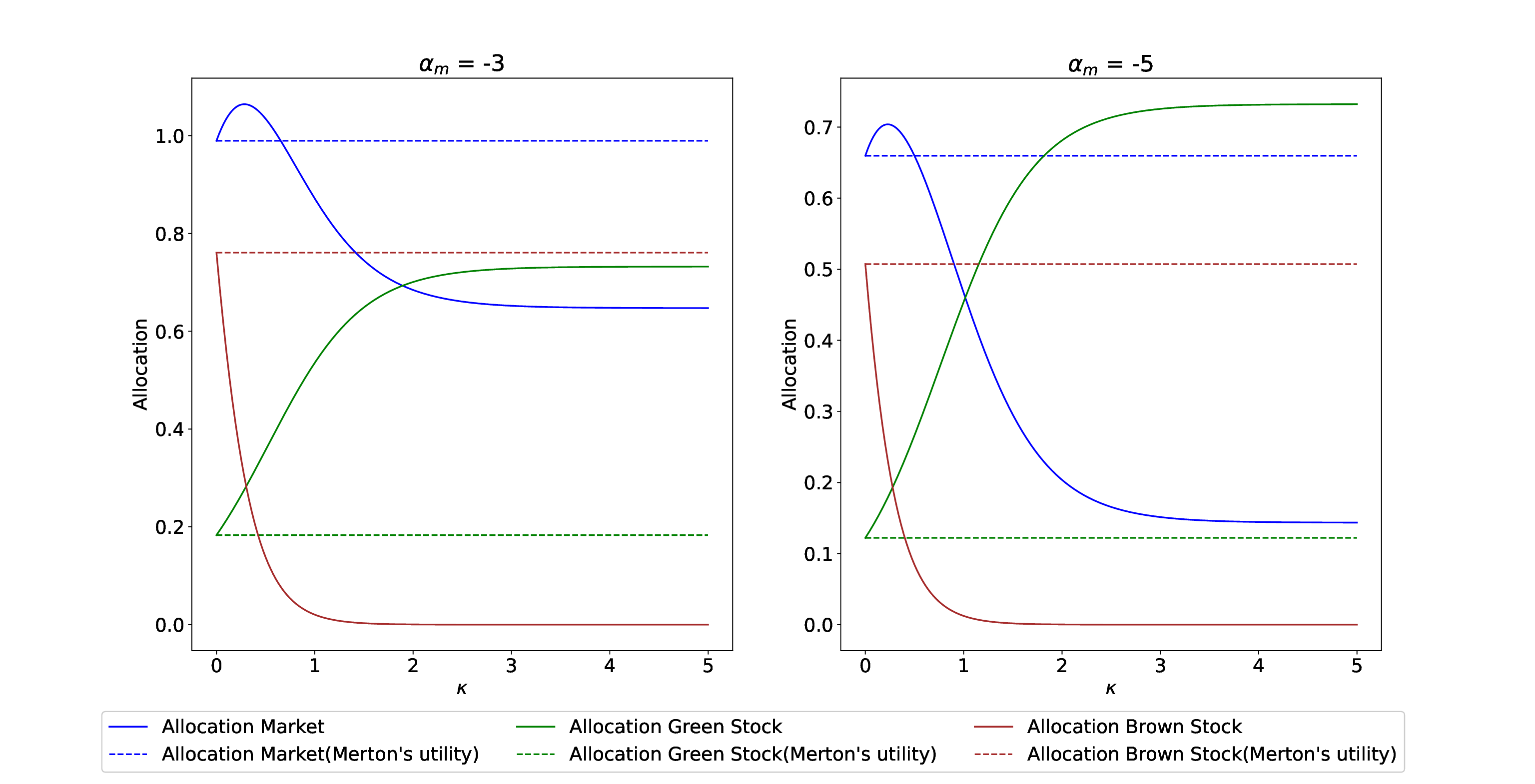"}}
	\caption{Impact of \(\kappa\) on allocation, IDT vs. WMT}
	\label{fig:kappa_alloc}
\end{figure}

To visually depict the impact of $\kappa$ on allocation weights, Figure \ref{fig:kappa_alloc} showcases the results for $\alpha_m = -3$ and $\alpha_m = -5$, focusing on the IDT and WMT context. In both cases, it is evident that as $\kappa$ increases from 0, there is an upward trend in the optimal weight assigned to the green stock, accompanied by a decrease in the weight allocated to the brown portfolio. Here we also see a non-constant behaviour on the allocation to the market portfolio, which shows an investor could favour both the market portfolio and the green stock at the expense of the brown. Moreover, the limiting scenario, as $\kappa$ goes to infinity, shows no brown investing while market and green investments persist. The deviation from the Merton solution at $\kappa=0$ aligns with the definition of $\kappa$ provided in our paper.

\begin{figure}
	\makebox[\textwidth][c]{\includegraphics[width=1.14\textwidth]{"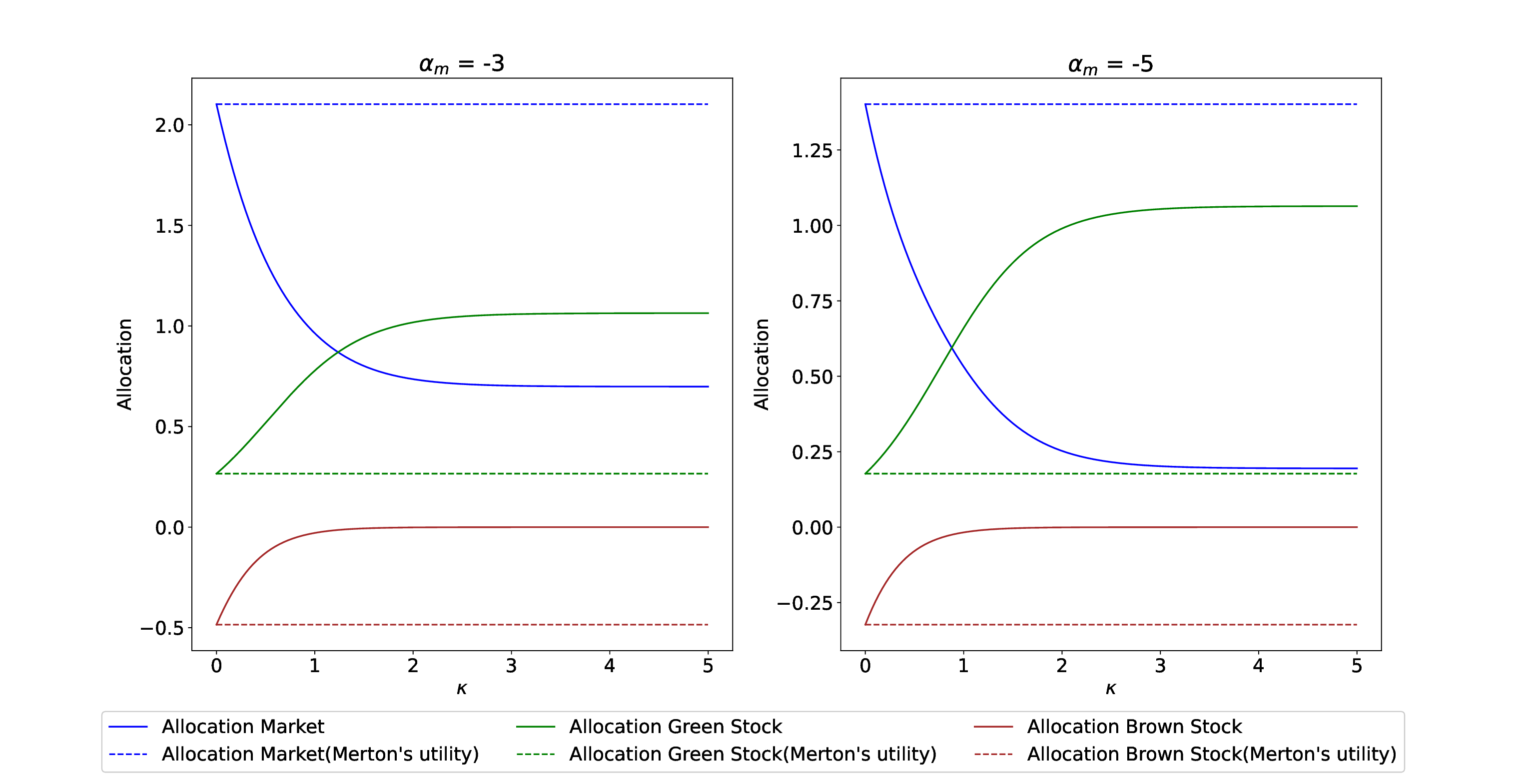"}}
	\caption{Impact of \(\kappa\) on allocation, Shenandoah vs. DuPont}
	\label{fig:kappa_alloc2}
\end{figure}

Moreover, when considering the Shenandoah and DuPont scenario, Figure \ref{fig:kappa_alloc2} demonstrates a similar pattern of increasing allocation toward the green stock as $\kappa$ rises. However, there is a notable difference compared with the findings presented in Figure \ref{fig:kappa_alloc}: the optimal weight for the brown stock also exhibits an apparent increase in value. The key is that this increase is a decrease in absolute value; in other words, the investor wants less risk coming from the brown investment. This contrast is not coincidental. As mentioned in subsection \ref{sec:impact}, when $\lambda_b < 0$, investors engage in short selling of the brown stock, and the extent of short selling decreases as $\kappa$ increases.

\subsection{GWEL and parameterization of ESG risk aversion levels}\label{sec:ea_gwel}

\begin{figure}
	\makebox[\textwidth][c]{\includegraphics[width=1.2\textwidth]{"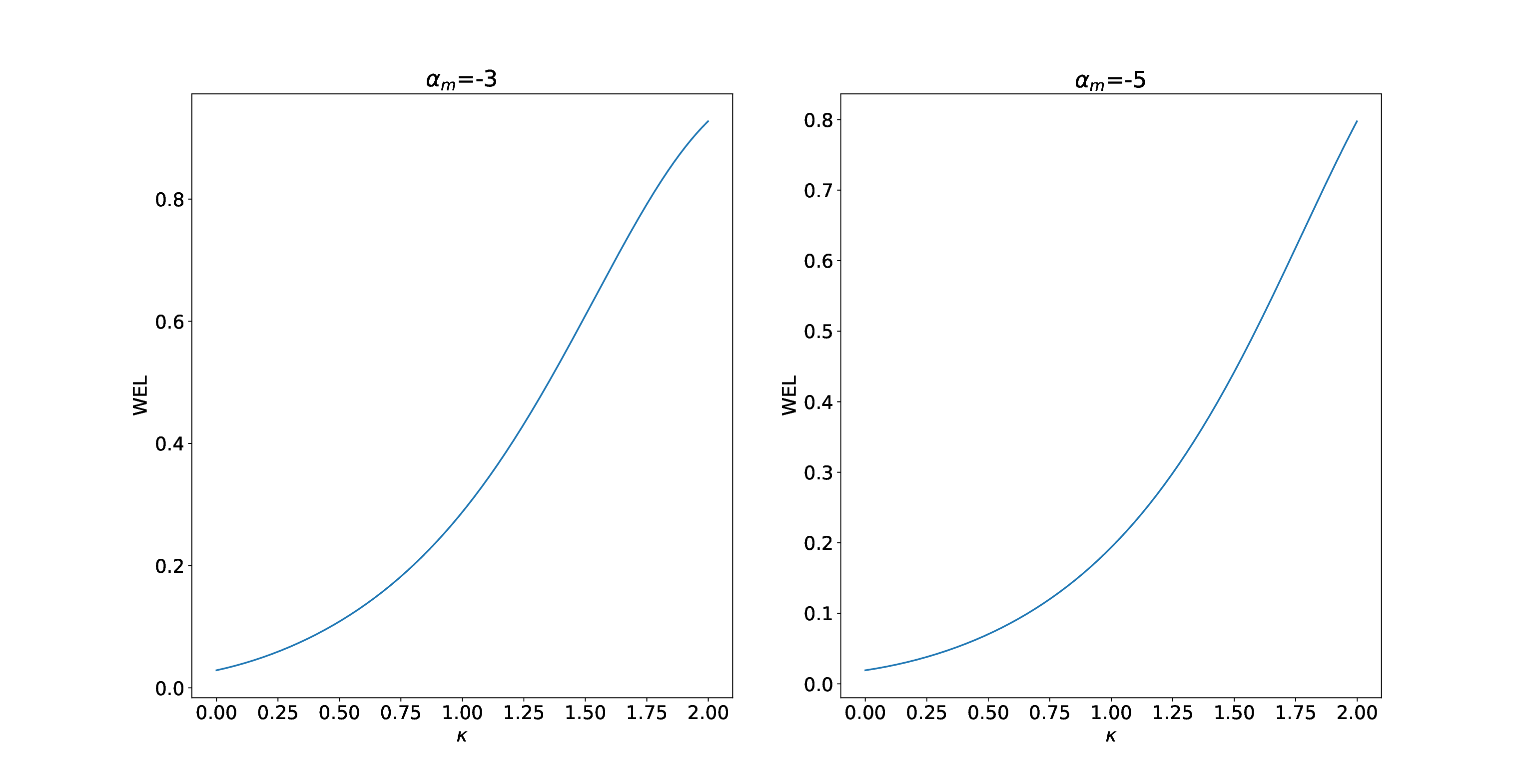"}}
	\caption{Impact of \(\kappa\) on GWEL employing Merton's solution, IDT vs. WMT}
	\label{fig:wel}
\end{figure}

\begin{figure}
	\makebox[\textwidth][c]{\includegraphics[width=1.2\textwidth]{"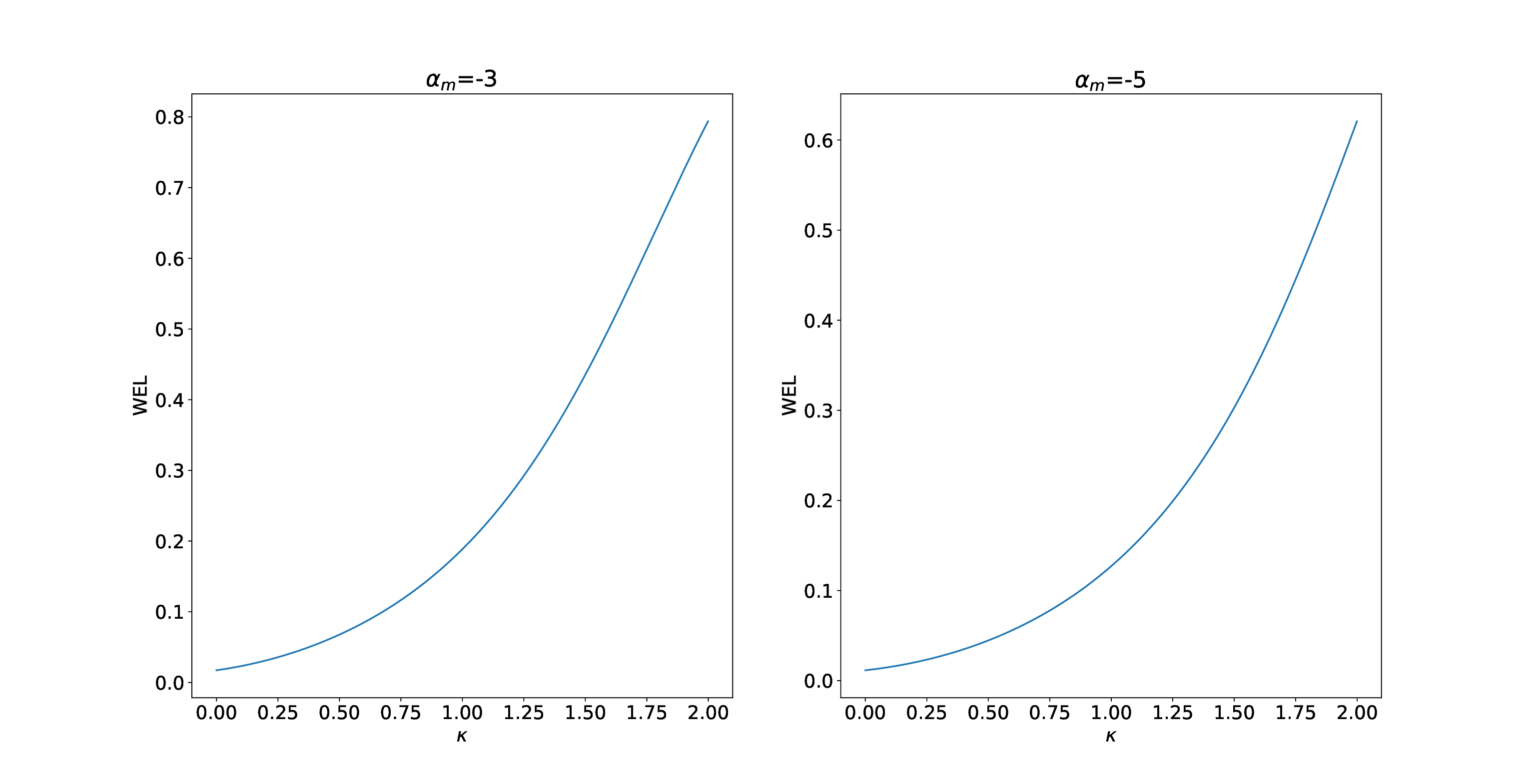"}}
	\caption{Impact of \(\kappa\) on GWEL employing Merton's solution, Shenandoah vs. DuPont}
	\label{fig:wel_2}
\end{figure}

In Section \ref{sec:3.3}, we compute the value of $q_m$ to investigate the GWEL when an investor employs the suboptimal strategy derived from Merton's solution. In order to avoid arbitrary choices for risk-aversion values, we work within the parameterization method outlined in Section \ref{sec:3.2}. Considering a one-year time horizon, we substitute $\alpha_i = \alpha_m \exp{(\kappa (E_m - E_i))}$, where $i \in {g, b}$, into equation (\ref{eqn:wel}), resulting in the generation of Figure \ref{fig:wel} and Figure \ref{fig:wel_2}. Both figures reveal a distinct pattern where the GWEL experiences a sharp increase as $\kappa$ rises. This indicates greater losses on Green Index for investors with stronger preferences for this index due to the use of the popular Merton's solution. As explained before, Merton's solution does not reflect the ESG preferences of the investor, leading to a larger GWEL as $\kappa$ increases.

In the IDT and WMT case, at $\kappa=2$, an investor would face a substantial wealth-equivalent loss of 92.73\% when $\alpha_m = -3$, and a loss of 79.76\% when $\alpha_m = -5$. As a by-product, these results suggest that the GWEL is higher for investors who are less risk-averse to the market. This can be explained by the fact that the lower the risk-aversion to the market portfolio, the lower the risk aversion to the green stock. Hence, investors who allocate more to the green stock would suffer the higher losses. The Shenandoah and DuPont case exhibits a similar trend, further confirming the significance of proper green allocation decisions.

\begin{figure}
	\makebox[\textwidth][c]{\includegraphics[width=1.2\textwidth]{"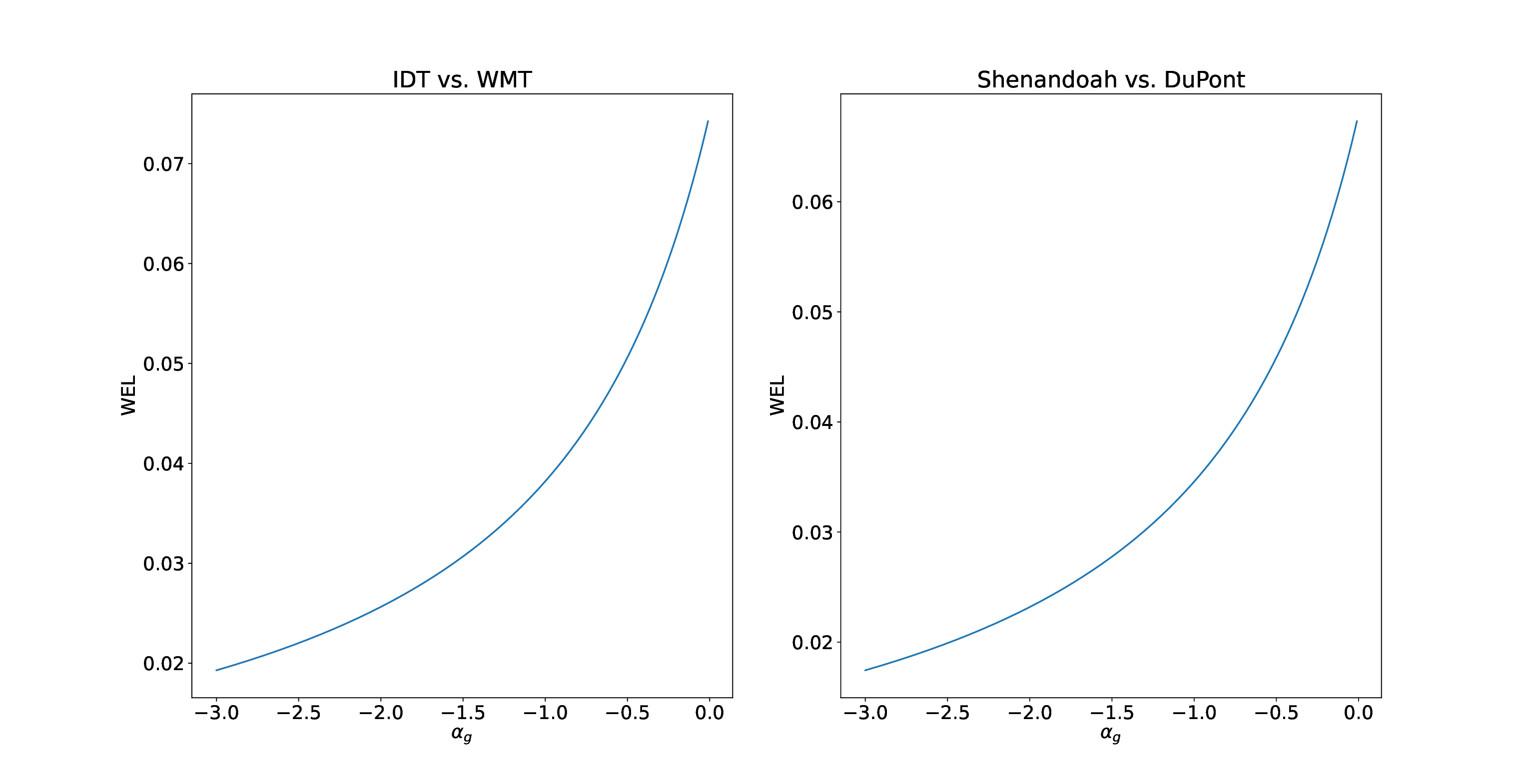"}}
	\caption{Impact of difference in green risk aversion on GWEL employing \(\pi_2=0\)}
	\label{fig:wel_3}
\end{figure}

Figure \ref{fig:wel_3} illustrates the GWEL $q_g$ for investors who choose not to invest in green assets. It is evident that $q_g$ consistently increases as investors become less risk averse toward the Green Index. In the IDT and WMT case, the GWEL reaches 7.42\% for $\alpha_g=-0.01$, while in the Shenandoah and DuPont case, the GWEL reaches 6.73\% for $\alpha_g=-0.01$.

These findings highlight the significant impact of suboptimal allocation strategies on the Green Index. Not only does such a strategy decrease the utility of the investor, but it also signifies a reduction in society's overall investment in green assets. This can hinder the development of ESG initiatives within the market, underscoring the importance of optimal allocation decisions.

\section{Conclusions}\label{sec:conclusion}

This research paper presents an expected multivariate utility analysis for ESG investors, employing a one-factor, CAPM-type structure to model green stocks, brown stocks, and a market index. The framework allows investors to adjust their risk aversion levels according to their preferences for green or brown investments. The study derives closed-form solutions for optimal allocations, wealth, and value functions. 

Notably, the findings demonstrate two significant outcomes. Firstly, they establish that investors can increase their green investments without compromising their overall pecuniary satisfaction. This highlights the potential for alignment between sustainability preferences and financial objectives. Secondly, the paper introduces a simple parameterization approach to capture investor preferences toward green assets, enhancing the model's applicability. 

Empirical analysis, based on RepRisk Ratings of U.S. stocks from 2010 to 2020, reveals substantial shifts in wealth allocation for ESG-sensitive investors, even while maintaining a constant level of pecuniary satisfaction. These insights contribute to a better understanding of how ESG considerations can be included in investment decisions, offering an encouraging conclusion as to the importance of increasing green investing to better accommodate our deep care for sustainable future.

\section*{Data availability statement}
The ESG ratings of the companies utilized in our empirical analysis were obtained from the RepRisk ESG Dataset. For additional information, please refer to the website: \url{https://www.reprisk.com/}.

\section*{Disclosure statement}
No potential conflict of interest was reported by the authors.

\bibliographystyle{tfcad}
\bibliography{Optimization_with_ESG_utility}

\clearpage

\appendix
\section{Proof of Proposition \ref{Prop1}}\label{apd:HJB}

\subsection{Derivation of the HJB equation}\label{apd:HJB_eqn}
Express equation (2.1) in matrix form, we have:
\begin{equation}
	d\bm{S_t} = \diag (\bm{S_t})[(r\bm{1} + \bm{\Omega}^{\frac{1}{2}}\bm{\lambda})dt+\bm{\Omega}^{\frac{1}{2}}d\bm{z_t}]
\end{equation}
where $\bm{S_t} = [S_{1,t}, S_{2,t}, S_{3,t}]^{\tran}$, $\bm{\lambda} = [\lambda_1\sigma_1, \lambda_g\sigma_2, \lambda_b\sigma_3]^{\tran}$, $\bm{z_t}=[z_{m,t}, z_{g,t}, z_{b,t}]^{\tran}$, and
\begin{equation}
	\bm{\Omega}^{\frac{1}{2}} = \begin{bmatrix}
		\sigma_1 & 0 & 0 \\
		\sigma_2\rho_{12} & \sigma_2\sqrt{1-\rho_{12}^2} & 0 \\
		\sigma_3\rho_{13} & 0 & \sigma_3\sqrt{1-\rho_{13}^2} 
	\end{bmatrix}
\end{equation}
Note that $\bm{\Omega}^{\frac{1}{2}} (\bm{\Omega}^{\frac{1}{2}})^{\tran}$ denotes the covariance matrix of $\bm{S_t}$.

Following equation (2.5), we have:
\begin{equation}
	\frac{dW_t}{W_t} = (r + \bm{\pi}^{\tran}\bm{\Omega}^{\frac{1}{2}}\bm{\lambda})dt + \bm{\pi}^{\tran}\bm{\Omega}^{\frac{1}{2}} d\bm{z_t}
\end{equation}
where $\bm{\pi} = [\pi_1, \pi_2, \pi_3]^{\tran}$. From equation (2.8):
\begin{equation}
	d\bm{X_t} = \diag (\bm{X_t})(\bm{r}dt + \bm{\Lambda} \bm{\lambda}dt+\bm{\Lambda}d\bm{z_t})
\end{equation}
where $\bm{X_t} = [X_{m,t}, X_{g,t}, X_{b,t}]^{\tran}$, $\bm{r} = [\theta_m, \theta_g, \theta_b]^{\tran}r$ and
\begin{equation}
	\bm{\Lambda} = \begin{bmatrix}
		\beta_p\sigma_1 & 0 & 0 \\
		0 & \pi_2\sigma_g & 0 \\
		0 & 0 & \pi_3\sigma_b 
	\end{bmatrix}
\end{equation}

Define the value function by
\begin{equation}
	J(\bm{X_t},t) = \sup_{\bm{\pi}} \expect_{\bm{X_t},t}[u(\bm{X_T})]
\end{equation}
with utility function $u(\bm{X_t})$ defined in equation (2.11) and the terminal condition
\begin{equation}
	J(\bm{X_T}, T) = u(\bm{X_T})
\end{equation}

Applying the tower property of conditional expectation on equation (2.21), we have:
\begin{align}
	\begin{split}
		J(\bm{X_t},t) &= \sup_{\bm{\pi}} \expect_{\bm{X_t},t}[\expect_{\bm{X_{t+\Delta t}},{t+\Delta t}}[u(\bm{X_T})]] \\
		& = \sup_{\bm{\pi}} \expect_{\bm{X_t},t}[J(\bm{X_{t+\Delta t}},{t+\Delta t})]
	\end{split}
\end{align}

Subtracting both sides by $J(\bm{X_t},t)$, we have:
\begin{equation}
	0=\sup_{\bm{\pi}} \expect_{\bm{X_t},t}[J(\bm{X_{t+\Delta t}},{t+\Delta t})-J(\bm{X_t},t)]
\end{equation}

Divide both sides by $\Delta_t$ then applying Ito's formula, we get:
\begin{equation}\label{eqn:apd.HJB}
	0 = \sup_{\bm{\pi}} \left\{ \nabla_t J + \nabla_{\bm{X}}J\diag (\bm{X_t}) (\bm{r} +\bm{\Lambda} \bm{\lambda}) + \frac{1}{2} \tr (\nabla^2_{\scaleto{\bm{X}\bm{X}}{4pt}}J \diag (\bm{X_t})^2 \bm{\Lambda}^2) \right\}
\end{equation}

\subsection{Derivation for the value function, optimal weights and optimal wealth}\label{apd:HJB_result}

Taking derivative of $\bm{\pi}$ on both sides of equation (\ref{eqn:apd.HJB}), and denoting $X_{m,t}$, $X_{g,t}$, $X_{b,t} $ by $x$, $y$, $z$, we get the first order conditions:
\begin{align}
	\begin{split}
		0&=\lambda_1\sigma_1^2xJ_x + \beta_p\sigma_1^2x^2J_{xx}\\
		0&=\lambda_1\sigma_1\sigma_2\rho_{12}xJ_x + \beta_p\sigma_1\sigma_2\rho_{12}x^2J_{xx} + \lambda_g\sigma_2\sigma_gyJ_y + \pi_2\sigma_g^2y^2J_{yy} \\
		0&=\lambda_1\sigma_1\sigma_3\rho_{13}xJ_x + \beta_p\sigma_1\sigma_3\rho_{13}x^2J_{xx} + \lambda_b\sigma_3\sigma_bzJ_z + \pi_3\sigma_b^2z^2J_{zz} \\
	\end{split}
\end{align}

Solving the equation above, we get:
\begin{align}\label{eqn:apd.weights}
	\begin{split}
		\pi_1 &= -\frac{\lambda_1J_x}{xJ_{xx}} + \frac{\lambda_g\sigma_2\rho_{12}J_y}{\sigma_1\sqrt{1-\rho_{12}^2}yJ_{yy}} + \frac{\lambda_b\sigma_3\rho_{13}J_z}{\sigma_1\sqrt{1-\rho_{13}^2}zJ_{zz}} \\
		\pi_2 &= -\frac{\lambda_gJ_y}{\sqrt{1-\rho_{12}^2}yJ_{yy}}\\
		\pi_3 &= -\frac{\lambda_bJ_z}{\sqrt{1-\rho_{13}^2}zJ_{zz}}\\
	\end{split}
\end{align}

Plugging the weights above back into equation (\ref{eqn:apd.HJB}), we have the non-linear partial differential equation:
\begin{equation}
	J_t - \frac{1}{2} \lambda_1^2\sigma_1^2\frac{J_x^2}{J_{xx}} - \frac{1}{2} \lambda_g^2\sigma_2^2\frac{J_y^2}{J_{yy}} - \frac{1}{2} \lambda_b^2\sigma_3^2\frac{J_z^2}{J_{zz}} + J_xx\theta_mr + J_yy\theta_gr+ J_zz\theta_br= 0
\end{equation}

Assuming $J$ has the form $J(x,y,z,t) = \frac{x^{\alpha_m}}{\alpha_m}\frac{y^{\alpha_g}}{\alpha_g}\frac{z^{\alpha_b}}{\alpha_b}v(t)$, then:
\begin{equation}
	v' + bv = 0
\end{equation}
where $b = \frac{1}{2}\lambda_1^2\sigma_1^2\frac{\alpha_m}{1-\alpha_m} + \frac{1}{2}\lambda_g^2\sigma_2^2\frac{\alpha_g}{1-\alpha_g} + \frac{1}{2}\lambda_b^2\sigma_3^2\frac{\alpha_b}{1-\alpha_b}+ (\theta_m\alpha_m+\theta_g\alpha_g+\theta_b\alpha_b)r$.

Solving the ordinary differential equation above, we have:
\begin{equation}
	J(X_m, X_g, X_b ,t) = \frac{X_m^{\alpha_m}}{\alpha_m}\frac{X_g^{\alpha_g}}{\alpha_g}\frac{X_b^{\alpha_b}}{\alpha_b} \exp (b(T-t))
\end{equation}

Substituting the optimal value function above back to the equation (\ref{eqn:apd.weights}), the optimal weights are therefore:
\begin{align}
	\begin{split}
		\pi_1^* &= \frac{\lambda_1}{1-\alpha_m} - \beta_2\pi_2^* - \beta_3\pi_3^*\\
		\pi_2^* &= \frac{\lambda_g}{\sqrt{1-\rho_{12}^2}(1-\alpha_g)}\\
		\pi_3^* &= \frac{\lambda_b}{\sqrt{1-\rho_{13}^2}(1-\alpha_b)}\\
	\end{split}
\end{align}

Taking the optimal weights above back to equation (\ref{eqn:wealth_process}), we have the optimal wealth process:
\begin{equation}
	\frac{dW_t}{W_t} = rdt+ \frac{\lambda_1}{1-\alpha_m}\frac{dS_{m,t}}{S_{m,t}} + \frac{\lambda_g}{\sqrt{1-\rho_{12}^2}(1-\alpha_g)}\frac{dS_{g,t}}{S_{g,t}} + \frac{\lambda_b}{\sqrt{1-\rho_{13}^2}(1-\alpha_b)}\frac{dS_{b,t}}{S_{b,t}}
\end{equation}

\section{Derivation of suboptimal strategies}\label{apd:sub}

\subsection{Derivation for the value function with fixed weights}

Assuming employing a suboptimal strategy with fixed weights $\bm{\pi} = [\pi_1, \pi_2, \pi_3]^{\tran}$, equation (\ref{eqn:hjb}) can be rewritten as:
\begin{align}\label{apd:hjb_fix}
	\begin{split}
		0 = &J_t + xJ_x\lambda_1\sigma_1^2\beta_p + \frac{1}{2}x^2J_{xx}\sigma_1^2\beta_p^{2}
		+ yJ_y\lambda_g\sigma_2\sigma_g\pi_2 + \frac{1}{2}y^2J_{yy}\sigma_g^2\pi_2^{2}\\
		&+ zJ_z\lambda_b\sigma_3\sigma_b\pi_3 + \frac{1}{2}z^2J_{zz}\sigma_b^2\pi_3^{2}
		+ (xJ_x\theta_m + yJ_y\theta_g + zJ_z\theta_b)r
	\end{split}
\end{align}
where $x,y,z$ denotes $X_{m},X_{g},X_{b}$, $\beta_p = \pi_1 + \frac{\sigma_2}{\sigma_1}\rho_{12}\pi_2 + \frac{\sigma_3}{\sigma_1}\rho_{13}\pi_3$.

Suppose that $J(x,y,z,t)$ has the form $\frac{x^{\alpha_m}y^{\alpha_g}z^{\alpha_b}}{\alpha_m\alpha_g\alpha_b}v(t)$, equation (\ref{apd:hjb_fix}) now becomes:
\begin{align}
	\begin{split}
		0 =& v' + [\alpha_m\lambda_1\beta_p\sigma_1^2 -\frac{\alpha_m(1-\alpha_m)}{2}\beta_p^{2}\sigma_1^2 +\alpha_g\lambda_g\pi_2\sigma_2\sigma_g -\frac{\alpha_g(1-\alpha_g)}{2}\pi_2^{2}\sigma_g^2 \\
		&+\alpha_b\lambda_b\pi_3\sigma_3\sigma_b -\frac{\alpha_b(1-\alpha_b)}{2}\pi_3^{2}\sigma_b^2 +(\alpha_m\theta_m+\alpha_g\theta_g+\alpha_b\theta_b)r]v
	\end{split}
\end{align}
Solving the differential equation above, we have:
\begin{equation}\label{apd:value_fix}
	J(x,y,z,t) = \frac{x^{\alpha_m}y^{\alpha_g}z^{\alpha_b}}{\alpha_m\alpha_g\alpha_b}\exp\{b^*(T-t)\}
\end{equation}
where $b^*=\alpha_m\lambda_1\beta_p\sigma_1^2 -\frac{\alpha_m(1-\alpha_m)}{2}\beta_p^{2}\sigma_1^2 +\alpha_g\lambda_g\pi_2\sigma_2\sigma_g -\frac{\alpha_g(1-\alpha_g)}{2}\pi_2^{2}\sigma_g^2+\alpha_b\lambda_b\pi_3\sigma_3\sigma_b -\frac{\alpha_b(1-\alpha_b)}{2}\pi_3^{2}\sigma_b^2 +(\alpha_m\theta_m+\alpha_g\theta_g+\alpha_b\theta_b)r$.

\subsection{Value function for the suboptimal strategy employing Merton's solution}\label{apd:sub_merton}

The optimal weights for the Merton's solution are:
\begin{align}
	\begin{split}
		\pi_1^* &= \frac{\lambda_1}{1-\alpha_m} - \frac{\sigma_2}{\sigma_1}\rho_{12} \pi_2^* - \frac{\sigma_3}{\sigma_1}\rho_{13} \pi_3^*\\
		\pi_2^* &= \frac{\lambda_g}{\sqrt{1-\rho_{12}^2}(1-\alpha_m)}\\
		\pi_3^* &= \frac{\lambda_b}{\sqrt{1-\rho_{13}^2}(1-\alpha_m)}\\
	\end{split}
\end{align}

Substituting the weights above to the equation (\ref{apd:value_fix}), we have:
\begin{align}
	\begin{split}
		b^* =& \frac{1}{2}\lambda_1^2\sigma_1^2\frac{\alpha_m}{1-\alpha_m}+\lambda_g^2\sigma_2^2\frac{\alpha_g}{1-\alpha_m}-\frac{1}{2}\lambda_g^2\sigma_2^2\frac{\alpha_g(1-\alpha_g)}{(1-\alpha_m)^2}
		+\lambda_b^2\sigma_3^2\frac{\alpha_b}{1-\alpha_m}\\
		&-\frac{1}{2}\lambda_b^2\sigma_3^2\frac{\alpha_b(1-\alpha_b)}{(1-\alpha_m)^2}+(\theta_m\alpha_m+\theta_g\alpha_g+\theta_b\alpha_b)r
	\end{split}
\end{align}

\subsection{Value function for the suboptimal strategy with $\pi_2=0$}\label{apd:sub_nongreen}

Rewriting equation (\ref{eqn:apd.HJB}) by specifying $\pi_2 = 0$:
\begin{align}
	\begin{split}
		0 = &\sup_{\bm{\pi}}\{J_t + xJ_x\lambda_1\sigma_1^2\beta_p + \frac{1}{2}x^2J_{xx}\sigma_1^2\beta_p^2 + zJ_z\lambda_b\sigma_3\sigma_b\pi_3 + \frac{1}{2}z^2J_{zz}\sigma_b^2\pi_3^2\\
		&+ (xJ_x\theta_m + yJ_y\theta_g + zJ_z\theta_b)r\}
	\end{split}
\end{align}
where $\beta_p = \pi_1 + \frac{\sigma_3}{\sigma_1}\rho_{13}\pi_3$.

The FOC would be:
\begin{align}
	\begin{split}
		0&=\lambda_1\sigma_1^2xJ_x + \beta_p\sigma_1^2x^2J_{xx}\\
		0&=\lambda_1\sigma_1\sigma_3\rho_{13}xJ_x + \beta_p\sigma_1\sigma_3\rho_{13}x^2J_{xx} + \lambda_b\sigma_3\sigma_bzJ_z + \pi_3\sigma_b^2z^2J_{zz} \\
	\end{split}
\end{align}
with the solution $\pi_1 = -\frac{\lambda_1J_x}{xJ_{xx}}-\frac{\sigma_3}{\sigma_1}\rho_{13}\pi_3, \pi_3 = -\frac{\lambda_b\sigma_3J_z}{\sigma_bzJ_{zz}}$.

Combining the two equations above, we have:
\begin{equation}
	0 = J_t  - \frac{1}{2}\lambda_1^2\sigma_1^2\frac{J_x^2}{J_{xx}} - \frac{1}{2}\lambda_b^2\sigma_3^2\frac{J_z^2}{J_{zz}} + (xJ_x\theta_m + yJ_y\theta_g + zJ_z\theta_b)r
\end{equation}

Assuming that $J(x,y,z,t)$ has the form $\frac{x^{\alpha_m}y^{\alpha_g}z^{\alpha_b}}{\alpha_m\alpha_g\alpha_b}v(t)$, the ODE is:
\begin{equation}
	0 = v' + [\frac{1}{2}\lambda_1^2\sigma_1^2\frac{\alpha_m}{1-\alpha_m}+\frac{1}{2}\lambda_b^2\sigma_3^2\frac{\alpha_b}{1-\alpha_b}+(\theta_m\alpha_m+\theta_g\alpha_g+\theta_b\alpha_b)r]v
\end{equation}

Solving the equation above, we obtain:
\begin{equation}
	J(x,y,z,t) = \frac{x^{\alpha_m}y^{\alpha_g}z^{\alpha_b}}{\alpha_m\alpha_g\alpha_b}\exp\{b(T-t)\}
\end{equation}
where $b=\frac{1}{2}\lambda_1^2\sigma_1^2\frac{\alpha_m}{1-\alpha_m}+\frac{1}{2}\lambda_b^2\sigma_3^2\frac{\alpha_b}{1-\alpha_b}+(\theta_m\alpha_m+\theta_g\alpha_g+\theta_b\alpha_b)r$.

The optimal weights are therefore:
\begin{align}
	\begin{split}
		\pi_1^* &= \frac{\lambda_1}{1-\alpha_m}  - \frac{\sigma_3}{\sigma_1}\rho_{13} \pi_3^*\\
		\pi_2^* &= 0\\
		\pi_3^* &= \frac{\lambda_b}{\sqrt{1-\rho_{13}^2}(1-\alpha_b)}\\
	\end{split}
\end{align}
Let us treat the optimal weights above as a suboptimal fixed strategy in equation (\ref{apd:value_fix}), we get:
\begin{equation}
	J(x,y,z,t) = \frac{x^{\alpha_m}}{\alpha_m}\frac{y^{\alpha_g}}{\alpha_g}\frac{z^{\alpha_b}}{\alpha_b} \exp (b^*(T-t))
\end{equation}
with $b^* = \frac{1}{2}\lambda_1^2\sigma_1^2\frac{\alpha_m}{1-\alpha_m} + \frac{1}{2}\lambda_b^2\sigma_3^2\frac{\alpha_b}{1-\alpha_b}+(\theta_m\alpha_m+\theta_g\alpha_g+\theta_b\alpha_b)r$.

\end{document}